\documentclass{aa}
\usepackage{graphicx}
\usepackage[varg]{txfonts}

\newcommand{\mics}{$\mu$m~}

\def\tex {\ifmmode{{T}_{\rm ex}}\else{$T_{\rm ex}$}\fi}
\def\tmb {\ifmmode{{T}_{\rm mb}}\else{$T_{\rm mb}$}\fi}
\def\ci     {\ifmmode{{\rm C}{\rm \small I}}\else{C\ts {\scriptsize I}}\fi}
\def\hi     {\ifmmode{{\rm H}{\rm \small I}}\else{H\ts {\scriptsize I}}\fi}
\def\hh     {\ifmmode{{\rm H}_2}\else{H$_2$}\fi}

\def\ts     {\thinspace}
\def\kms    {\ifmmode{{\rm \ts km\ts s}^{-1}}\else{\ts km\ts s$^{-1}$}\fi}
\def\msol   {\ifmmode{{\rm M}_{\odot}}\else{M$_{\odot}$}\fi}
\def\lsol   {\ifmmode{{\rm L}_{\odot}}\else{L$_{\odot}$}\fi}
\def\zsol   {\ifmmode{{\rm Z}_{\odot}}\else{Z$_{\odot}$}\fi}
\def\etal   {{\rm et\ts al.}~}

\begin{document}

\title{ALMA observations of molecular tori around massive black holes
\thanks{Based on observations carried out with  ALMA in cycles 3 and 4.
}}

\author{F. Combes \inst{1,2}
\and
S. Garc\'{\i}a-Burillo \inst{3}
\and
A. Audibert \inst{1}
\and
L. Hunt \inst{4}
\and
A. Eckart \inst{5}
\and
S. Aalto \inst{6}
\and
V. Casasola \inst{4,7}
\and
F. Boone \inst{8}
\and
M. Krips \inst{9}
\and
S. Viti \inst{10}
\and
K. Sakamoto \inst{11}
\and
S. Muller \inst{6}
\and
K. Dasyra \inst{12}
\and
P. van der Werf \inst{13}
\and
S. Martin \inst{14,15}
           }
\institute{Observatoire de Paris, LERMA, CNRS, PSL Univ., Sorbonne University, UPMC, Paris, France
 \and  
 College de France, 11 Pl. Marcelin Berthelot, 75231, Paris
 \and 
Observatorio Astron\'omico Nacional (OAN)-Observatorio de Madrid,
Alfonso XII, 3, 28014-Madrid, Spain
 \and 
INAF - Osservatorio Astrofisico di Arcetri, Largo E. Fermi, 5, 50125, Firenze, Italy
 \and 
I. Physikalisches Institut, Universit\"at zu K\"oln, Z\"ulpicher Str. 77, 50937, K\"oln, Germany
 \and  
 Dep. of Space, Earth and Environment, Chalmers University of Technology, Onsala Space Observatory, SE-43992 Onsala, Sweden
 \and 
INAF -- Istituto di Radioastronomia,  via Piero Gobetti 101, 40129, Bologna, Italy
\and  
CNRS, IRAP, 9 Av. colonel Roche, BP 44346, 31028, Toulouse Cedex 4, France
 \and 
RAM, 300 rue de la Piscine, Domaine Universitaire, F-38406 Saint Martin d'H\`eres, France
 \and  
 Dep. of Physics and Astronomy, UCL, Gower Place, London WC1E 6BT, UK 
 \and 
Academia Sinica, Institute of Astronomy and Astrophysics,  Taiwan
 \and 
 Dep. of Astrophysics, Astronomy \& Mechanics, Faculty of Physics, National and Kapodistrian University of Athens, Panepistimiopolis Zografou, 15784, Greece, and  National Observatory of Athens, Institute for Astronomy, Astrophysics, Space Applications and Remote Sensing, Penteli, 15236, Athens, Greece
 \and 
 Leiden Observatory, Leiden Univ., PO Box 9513, 2300 RA Leiden, Netherlands
 \and 
 European Southern Observatory, Alonso de C\'ordova, 3107, Vitacura, Santiago 763-0355, Chile
 \and 
 Joint ALMA Observatory, Alonso de C\'ordova, 3107, Vitacura, Santiago 763-0355, Chile
              }

   \date{Received  2017/ Accepted  2017}

   \titlerunning{Molecular tori}
   \authorrunning{F. Combes et al.}

   \abstract{We report Atacama Large Millimeter/submillimeter Array (ALMA) observations
     of CO(3-2) emission in a sample of seven Seyfert/LINER galaxies
 at the unprecedented spatial resolution of 0\farcs1\,=\,4-9~pc. 
 Our aim is to explore the close environment of  active galactic nuclei (AGN), and the dynamical 
 structures leading to their fueling, through the morphology and
 kinematics of the gas inside the sphere of influence of the black hole.  
 The selected galaxies host low-luminosity AGN and have a wide range of activity types 
 (Seyferts 1 to 2, LINERs), and barred or ringed morphologies.
  The observed maps reveal the existence of circumnuclear disk structures, defined by
  their morphology and decoupled kinematics, in most of the sample. We
  call these structures molecular tori, even though they often appear  as disks 
  without holes in the center. They have
   varying orientations along the line of sight, unaligned with the host galaxy 
   orientation. The radius of the tori ranges
  from 6 to 27 pc, and their mass from 0.7 $\times$ 10$^7$ to 3.9 $\times$ 10$^7$ M$_\odot$.
 The most edge-on orientations of the torus correspond to obscured Seyferts.
  In only one case (NGC~1365), the AGN is centered on the central gas hole of the torus. 
  On a  larger scale, the gas is always piled up in a few  resonant rings 100~pc in scale 
  that play the role of a reservoir to fuel the nucleus. In some cases, a trailing 
  spiral is observed inside the ring, providing evidence
 for feeding processes. More frequently, the torus and the AGN are slightly
 off-centered with respect to the bar-resonant ring position, implying that the black hole
 is wandering by a few 10~pc amplitude around the center of mass of the galaxy.
 Our spatial resolution allows us to measure gas velocities inside the sphere of
 influence of the central black holes.
 By fitting the observations with different simulated cubes, varying the torus 
 inclination and the black hole mass,
  it is possible to estimate the mass of the central
 black hole, which is in general difficult for such late-type galaxies, with only a pseudo-bulge.
  In some cases, AGN feedback is revealed through a molecular outflow, which will be studied
 in detail in a subsequent article. }


\maketitle


\section{Introduction}
\label{intro}

The growth of supermassive black holes in galaxies produces phenomenon of  active galactic nuclei (AGN), 
one of the brightest and most energetic events in the Universe. In recent years, the subsequent appearance of
AGN feedback has been widely established through the existence of fast outflows of ionized
and atomic gas  \citep[e.g.,][]{Veilleux2005, Tombesi2010, Fiore2017}.
In parallel, observations of the molecular 
component of the circumnuclear environment have brought a great deal of  progress in the question of how AGN are 
fueled in galaxies \citep[e.g.,][]{Garcia-Burillo2005, Combes2013, Combes2014}, and how the energy generated by the AGN can in turn regulate its gas accretion 
through molecular outflows \citep[e.g.,][]{Feruglio2010, Aalto2012, Cicone2014, Garcia-Burillo2014}. This has important implications 
for the co-evolution of galaxies and black holes which is observed through the now well-established M$_{\rm BH}$-$\sigma$
relation \citep[e.g.,][]{Gultekin2009}.  

Active galactic nuclei are observed in two categories, type 1 with broad-line regions (BLR) and type 2 with only narrow-line 
regions (NLR). Lines are broad only very close to the black hole, in the accretion disk, while they are narrow 
farther out in the NLR of $\sim$0.1-1~kpc size. 
The original unification paradigm proposes that the BLR in type 2 is 
obscured by a dusty molecular torus, along the line of sight of the observer 
\citep[e.g.,][]{Antonucci1985, Urry1995}. However, since this early work many observations have shown
that inclination and obscuration are not the only parameters distinguishing types 1 and 2; some
of these types are intrinsically different and/or nuclear starbursts are confusing the picture
 \citep[e.g.,][]{Imanishi2004, Hatziminaoglou2009}. A strong challenge of the unification paradigm
  also comes from
AGN changing look from type 1 to type 2 and vice versa, without evidence
of variable obscuration \citep[e.g.,][]{LaMassa2015, McElroy2016}.

The expected torus is so small (3-30~pc in size) that it was not possible to  
resolve it until recently, where CO(6-5) emission was detected for 
the first time with ALMA, as well as continuum and dense gas tracers,
in a 10~pc-diameter torus in the Seyfert 2 NGC 1068 
\citep{Garcia-Burillo2016, Gallimore2016, Imanishi2016, Imanishi2018}.
Dusty tori have also been seen in the near or mid-infrared \citep[e.g.,][]{Asmus2011, Gratadour2015},
although sometimes the dust emission is seen in the polar direction instead \citep{Asmus2016}.

\cite{Garcia-Burillo2016} have obtained a high-resolution map in CO(6-5) with ALMA towards the 
center of NGC 1068: there is a circumnuclear disk (CND) 
 $\sim$300~pc in size, which is also detected in dust continuum. The AGN is offset 
with respect to the center of this disk. 
Around the AGN, a peak of CO emission is detected. This is identified 
as the molecular torus surrounding the AGN. 
The dust emission coincides spatially with the molecular torus. Two components were revealed, 
 a dust torus  $\sim$7~pc in diameter, oriented along PA = 142$^\circ$, aligned with the H$_2$O maser disk 
\citep{Greenhill1996} and some polar emission that extends 10~pc to the SW. The CO(6-5) torus is a 
bit larger in size (diameter $\sim$ 10~pc) than the dust torus, and it appears tilted, 
with a PA = 112$^\circ$
 with respect to the dust torus and accretion disk. The extent of the torus depends slightly
 on its tracer; it is somewhat larger with low-J CO emission \citep{Garcia-Burillo2018}
 or in dense gas tracers \citep{Imanishi2018}.
 There is no CO counterpart for the polar emission. 
The molecular torus reveals strong non-circular motions and a large degree of turbulence. It also appears  
more face-on at larger radii, being probably warped. The perturbations in the morphology 
and kinematics of the torus can be  
interpreted in terms of the \cite{Papaloizou1984} instability (PPI), predicted in particular for the 
dynamical evolution of AGN tori.

The discovery of such a perturbed and turbulent torus was a surprise. 
Is it due to continuous accretion from
 the CND onto the torus, triggering tilt, warp, and PPI instabilities, 
 and finally leading to the AGN fueling? 
 How does the molecular outflow detected farther away \citep{Garcia-Burillo2014} arise from the torus? 
Are perturbed and turbulent tori the norm in AGN-dominated galaxies?

In the present paper, we describe our effort to gather more information on possible molecular tori
in nearby Seyfert galaxies. We have observed with high spatial resolution 
seven barred galaxies with active nuclei
in order to explore both the AGN fueling and the feedback processes, and also to characterize the molecular
content inside the central kpc at 4-9 pc resolution. In many cases the gas 
kinematics allows us to refine the
determination of the central black hole mass. Section \ref{sample} presents 
the galaxy sample, and  Section \ref{obs}
the details of the ALMA observations. The results are then described in Section \ref{res} with first a
brief discussion of the main features discovered, whether there is  a molecular torus, the determination
of its properties, and then the estimation of black hole masses.
Section \ref{disc} summarizes and discusses our findings.

\begin{figure}[ht]
\centerline{
\includegraphics[angle=0,width=8.5cm]{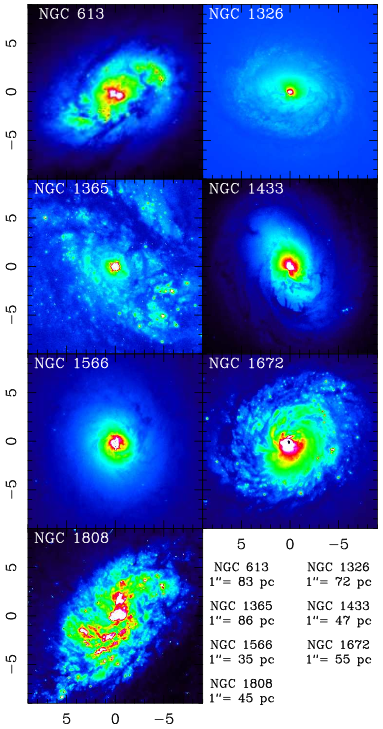}
}
\caption{ Red HST images (F814W) for the seven sample galaxies in the same FOV = 18 \arcsec\ as
  obtained with ALMA in Band 7. The axis labels correspond to arcsec, with north up and east to the left. 
}
\label{fig:HST}
\end{figure}

\section{The sample}
\label{sample}
In addition to NGC~1433 and 1566 \citep{Combes2013, Combes2014},
we  selected  the five nearby southern AGN galaxies for which 
we previously gathered  CO(3-2) observations at 0.14’’ resolution. They span more than a factor of 100 in AGN 
power (X-ray and radio luminosities), a factor of 10 in star formation 
rate (SFR), and a wide range of galaxy 
inner morphology (with or without double bars). This sample has been selected to
provide a wide range of gas inflow rate, 
AGN feeding rate, and therefore test the various possible phases of evolution for the molecular tori.  
The galaxies were also selected to be sufficiently nearby to allow ALMA, with its exquisite spatial resolution, to resolve the torus if present. 
All galaxies have single-dish millimeter data with the SEST
(see references in Table \ref{tab:sample}), and we have 
obtained ALMA cycle 0 or 3 intermediate resolution data, in addition to the 
present high-resolution ALMA cycle 3-4 CO maps. 
Some of the targets have
also been observed  by previous authors with ALMA at lower resolution of 50-100~pc
\citep[e.g., for NGC~613 and NGC~1808][]{Miyamoto2017, Salak2016, Salak2017}. 
We  used these observations in the archive when available. Our targets have high-resolution Hubble Space Telescope (HST)
images, and are found in the IRAS Bright Galaxy Sample \citep{Sanders2003}. 
Most of these galaxies have been searched 
for H$_2$O masers (tracing the accretion disk) with a 2/7 detection rate \citep{Zhang2012, Surcis2009} 
and for star-forming (``active'') inner rings by Comeron (2013). In addition, we have proprietary SINFONI IFU 
observations of all galaxies, so that we can compare warm H$_2$ morphology, ionized gas distributions, and 
kinematics with the cold molecular counterparts. We summarize the galaxy 
properties in Table \ref{tab:sample}, and the 
nuclear morphologies in Figure \ref{fig:HST}.

\begin{table*}[h!]
      \caption[]{Characteristics of the sample}
         \label{tab:sample}
            \begin{tabular}{l c c c c c c c c c c }
            \hline
            \noalign{\smallskip}
   Name&   Type &    D   &SFR    & log(L$_X$)   & log(L$_{1.4GHz}$)& S(CO)$_{21}$ & Bar  & Double &RA & Dec \\
            \noalign{\smallskip}
       &        &  Mpc & M$_\odot$/yr& erg/s & W/Hz & Jy km/s& PA($^\circ$)   & bar? & ICRS & ICRS \\
            \hline
            \noalign{\smallskip}
N613 & Sy- SB(rs)bc  & 17.2& 5.3$^*$& 41.2& 21.8 &1200$^{\mathrm{a}}$ & 127 & 122 & 01:34:18.189&  -29:25:06.59\\
N1326& LINER- SB0(r)  & 14.9& 1.1& 39.9& 20.9  & 200$^{\mathrm{b}}$ &  30  &  90& 03:23:56.416 &-36:27:52.68\\
N1365& Sy 1.8- SB(s)b & 17.8& 17.& 41.8  & 22.3 & 2300$^{\mathrm{c}}$  &  92  &  46 &03:33:36.368 &-36:08:25.51\\
N1433& Sy 2- SB(r)ab  & 9.7& 0.5    & 39.2  & 20.0 & 360$^{\mathrm{a}}$  &  95  &  32&03:42:01.49 & -47:13:20.2\\
N1566& Sy 1.5- SAB(s)bc& 7.2& 0.8  & 40.5 & 21.3 & 540$^{\mathrm{a}}$  &  5  &  No&04:20:00.395&-54:56:16.61\\
N1672& Sy 2- SB(s)b  & 11.4& 3.1  & 38.4  & 19.9 & 1140$^{\mathrm{a}}$  &  97  &  No& 04:45:42.496 &-59:14:49.91\\
N1808& Sy 2- SAB(s)a & 9.3& 4.7   & 39.8   & 21.7 & 4500$^{\mathrm{d}}$  & 139  &  158&05:07:42.329&-37:30:45.85\\
            \noalign{\smallskip}
            \hline
            \end{tabular}
            \begin{list}{}{}
\item -- D are the median values of z-independent distances from NED \citep{Steer2017};\\
  -- SFR are derived from infrared luminosities (NED);\\
  -- $^*$ NGC~613 has an H$_2$O maser;\\
 -- L$_X$ is from 2-10keV INTEGRAL, Rosat, and/or Chandra archives;\\
 -- The CO(2-1) integrated fluxes are from single-dish measurements (SEST, beam 22\arcsec) from
  [$^{\mathrm{a}}$] \cite{Bajaja1995},  [$^{\mathrm{b}}$] \cite{Garcia-Barreto1991},
  [$^{\mathrm{c}}$] \cite{Sandqvist1999},  [$^{\mathrm{d}}$] \cite{Aalto1994};\\
  -- PA of bars are from \cite{Jungwiert1997} for NGC~613 and 1433, from \cite{Garcia-Barreto1991} for
  NGC~1326, \cite{Lindblad1999} for NGC~1365, \cite{Aguero2004} for NGC~1566, \cite{Jenkins2011} for
  NGC~1672, and \cite{Dahlem1994} for NGC~1808;\\
  -- The RA-Dec positions are the new adopted centers for each galaxy, derived from the detected continuum point sources in the present work, with an error bar of $\sim$ 0.1\arcsec (Sec. \ref{continuum}).
\end{list}
\end{table*}

%
\begin{center}
\begin{table}
  \caption[]{Configuration of the observations
        and resulting spatial resolution (with robustness parameter of 0.5, see Sect. \ref{obs})} 
\label{tab:config}
\begin{tabular}{lll}
\hline
Galaxy  & Cycle 3, TC \& TE & Cycle 4, TM2 \& TM1\\
\hline
N 613        & 0.41x0.35 \& 0.15x0.13\arcsec    & 0.33x0.30 \& .092x.064\arcsec  \\
N1326     & 0.35x0.26  \& 0.16x0.14\arcsec   & 0.36x0.29 \&  .086x.058\arcsec \\
N1365     & 0.33x0.24  \& 0.15x0.14\arcsec   & 0.35x0.28 \& .087x.060\arcsec  \\
N1433     & 0.56x0.42\arcsec$^{\mathrm{a}}$  & 0.37x0.30 \& .067x.059\arcsec  \\
N1566     & 0.64x0.43\arcsec$^{\mathrm{a}}$  & 0.35x0.29 \& .061x.045\arcsec  \\
N1672     & 0.35x0.19 \& 0.16x0.12\arcsec    & 0.41x0.29 \& .096x.064\arcsec  \\
N1808     & 0.36x0.24 \& 0.14x0.11\arcsec    & 0.30x0.29 \& .082x.064\arcsec  \\
\hline
\end{tabular}
\begin{list}{}{}
\item[-- $^{\mathrm{a}}$]Obtained in cycle 0 (Combes \etal 2013, 2014);\\
-- TC and TM2 are compact configurations, TE and TM1 are extended.
\end{list}
\end{table}
\end{center}

\section{Observations}
\label{obs}

To explore and characterize molecular tori in nearby Seyferts, we targeted the CO(3-2) line.
For this transition, the J is low enough that it is still a tracer of density and mass more than excitation;
it also has a relatively high flux and affords a high spatial resolution. 
 It appears as the best compromise
between spatial resolution, sensitivity, and field of view. 
To estimate H$_2$ column densities and masses,
we  use in Sect. \ref{tori} the usual ratios applicable to active galaxy centers.
The detailed study by \cite{Papadopoulos2012} has shown that the excitation ratios
begin to depart from common ones at $J=5$ and beyond, and mainly for starbursts.

The observations were carried out with the ALMA telescope  in cycles 3 and 4,
with 36 to 40 antennas, during the years 2016 and  2017. The corresponding ALMA
projects ID were \#2015.1.00404.S and \#2016.1.00296.S, both with PI F. Combes. 
In cycle 3, five galaxies (NGC~613, NGC~1326, NGC~1365, NGC~1672, and NGC~1808) were observed 
simultaneously in CO(3-2), HCO$^+$(4-3), HCN(4-3), and continuum, with Band 7.
The  compact configuration (TC, baselines 15 to 630m) and extended (TE, baselines
15 to 1400m) combined to give a synthesized beam
of 0\farcs14  ($\sim$ 15~pc), and an rms sensitivity of 1mJy/beam in 10km/s channels
(80 $\mu$Jy/beam in the continuum). The total integration
time, including calibration and overheads, was 1~h per source.
This choice of correlator configuration, selected to simultaneously observe 
three lines, provided  
a velocity range of 1600~km/s for each line. 
However, the various lines are not centered, and in particular
a compromise had to be made for the CO(3-2) and HCN(4-3). These two
lines are sampled with only
200~km/s on one side and 1400~km/s on the other. This is adequate
for nearly face-on galaxies, but prevents  seeing outflows on one side.
The bandwidth was 1800~MHz for the continuum bands.

In cycle 4, the seven galaxies were observed at higher spatial resolution,
$\sim$ 0\farcs07 or 4 to 9~pc (depending on  the various distances), to search for molecular tori.
For the frequency tuning we chose to observe the CO(3-2) and the HCO$^+$(4-3) lines, 
and the continuum in Band 7. The HCN(4-3) transition was not observed
in order to avoid a restricted velocity range in the expected broader spectral lines towards the nuclei.
The observations were done in several blocks, a compact configuration 
(TM2, baselines 19 to 500~m ) and extended
(TM1, baselines 19 to 3100~m), with a total duration of two hours per galaxy. 
When combining two or more of these configurations, all calibrated observations
with all baselines were included to obtain the UV-tables in CASA or GILDAS.
The sensitivity reached was
between 0.6 and 0.8mJy in 10~km/s channels. A summary of all configurations observed is given in Table
\ref{tab:config}.

The observations were all centered on the nuclei, with a single pointing covering a field of view (FOV)
of 18\arcsec. For NGC~1365, due to the large size of the galaxy and nuclear ring and the strength of CO emission, we performed a rectangular 13-point mosaic, ensuring a FOV of 45 $\times$ 36\arcsec, aligned on the major axis. The galaxies were observed
in dual polarization mode with 1.875~GHz total bandwidth per baseband, and
a channel spacing of 0.488 MHz corresponding to $\sim$0.8 km/s, after Hanning smoothing.
The spectra were then smoothed
to 11.7~MHz (10.2~km/s) to build channel maps.

The total integration time provided an rms of 30 $\mu$Jy/beam in the continuum,
and $\sim$0.6 mJy/beam in the line channel maps (corresponding to $\sim$1.2K
at the obtained spatial resolution).
The flux calibration was done with nearby quasars, which are
regularly monitored at ALMA, and resulted in 10\% accuracy.

The  data were  calibrated, imaged, and cleaned with the CASA software
(versions  4.5.3 to 4.7.2; \citealt{McMullin2007}) and the analysis
was then finalized with the GILDAS software \citep{Guilloteau2000}.
The final cubes at high resolution are at the maximum
1800x1800 pixels with 0\farcs01 per pixel in the plane
of the sky, and have 60 channels of 10~km/s width.
The maps were made with Briggs weighting and a robustness parameter of 0.5, i.e., 
a trade-off between uniform and natural weighting.
The data were cleaned using a mask made from the integrated CO(3-2) map.
The continuum was subtracted from all line maps.

The final maps were corrected for primary beam attenuation.
Very little CO(3-2) emission was detected outside the full width half power (FWHP) primary beam. 
Because of missing short spacings, the interferometer plays the role
of a high-frequency filter, insensitive to smooth and extended emission;
the scales that might be filtered out are those larger than $\sim$3\arcsec\, in each 
channel map. Since the velocity gradients are high in galaxy nuclei,
this does not affect  the line measurements too much; indeed,
the size in each velocity channel is not expected to be extended.
However, the missing-flux problem might affect continuum maps.
Low-level negative sidelobes adjacent to bright emission were
observed.

The exact values of missing flux amount will be detailed in forthcoming papers. 
We  estimated
the total CO(3-2) line flux for the two galaxies previously observed in cycle 0 (with the most compact configuration) to give
an order of magnitude (see Table \ref{tab:short}).
The difference between the cycle 0 observations with beams of 0\farcs5 and
the combination TM1+TM2 shows that we are missing $\sim$ 25\% of the total flux, 
on large angular scales. This will 
affect only slightly the masses estimated for the tori since they are not extended,
except for NGC~1566. In future papers, all configurations will be combined
to minimize flux losses.

%
\begin{center}
\begin{table}
  \caption[]{Estimation of the missing CO(3-2) flux in the FOV of 18\arcsec
    for the two galaxies observed in cycle 0} 
\label{tab:short}
\begin{tabular}{lccc}
\hline
Galaxy  & Cycle 0 &  TM2+TM1 & TM1\\
\hline
NGC~1433     & 234  & 174  & 161 \\
NGC~1566     & 596   &  402  &  364  \\
\hline
\end{tabular}
\begin{list}{}{}
\item[-- All integrated fluxes are in Jy km/s]
\end{list}
\end{table}
\end{center}

\begin{figure*}[h!]
\centerline{
\begin{tabular}{c}
  \includegraphics[angle=-0,width=17.5cm]{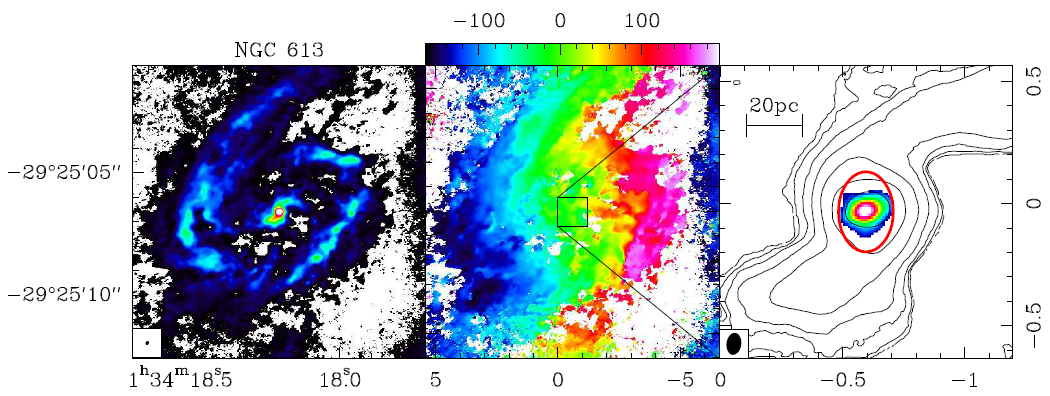}\\
\includegraphics[angle=-90,width=17.5cm]{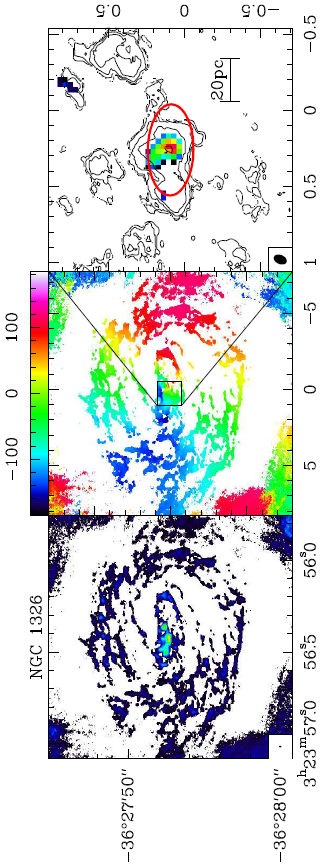}\\
\end{tabular}
}
\caption{{\it Top}: CO(3-2) moment 0 and 1 of galaxy NGC~613 (left and middle,  TE+TM1).
  The right panel is a zoom-in of the CO(3-2) contours (of the TM1 observations
  at high resolution) superposed on a color scale of the continuum emission in Band 7.
  The three panels have been centered on the central continuum source, and the labels
  are the offset in arcseconds from the phase center.
  For the two first panels the beam size is indicated at the bottom left.
  In the third panel is shown the ellipse (in red) used to define the molecular
  tori in Section \ref{tori}. The contours are from 25$\sigma$ to
  1600$\sigma$, and follow each other by factor 2 multiplication. The scale bar in the right panel is 20~pc long.
  The color scale in the middle panel is in km/s.
  {\it Bottom:} Same, but for galaxy NGC~1326. The contours are from 3$\sigma$ to
  96$\sigma$, and follow each other by factor 2 multiplication. 
}
\label{fig:613-1326}
\end{figure*}

\begin{figure*}[h!]
\centerline{
\begin{tabular}{c}
  \includegraphics[angle=-90,width=17.5cm]{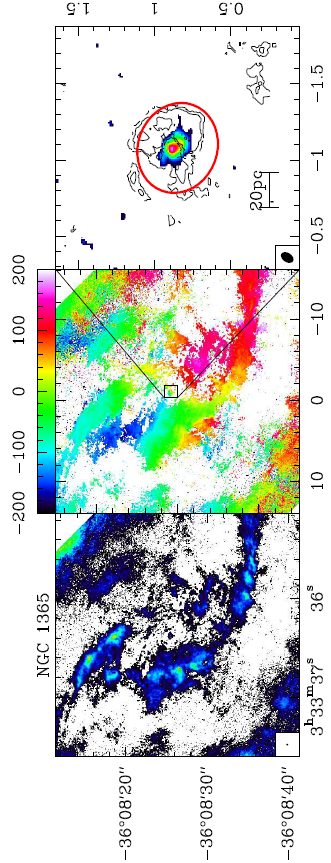}\\
\includegraphics[angle=-90,width=17.5cm]{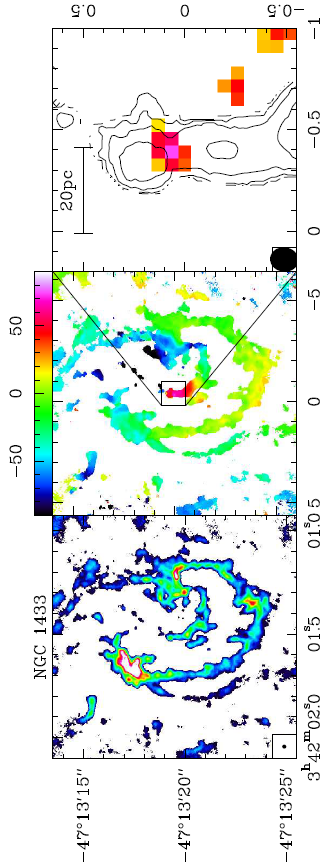}\\
\end{tabular}
}
\caption{{\it Top}: Same as fig \ref{fig:613-1326} for NGC~1365 (from the center of the mosaic;
  the size of the beam has been
  multiplied by 1.5 to become visible in the first panel) -- the contours are from 4$\sigma$ to
  256$\sigma$, and follow each other by factor 2 multiplication -- and   {\it Bottom:}
  for NGC~1433 (in the latter case,  TM2+TM1 in all 3 panels).  The continuum emission
  (color plot in the right panel)
  maximizes at 0.2 mJy (i.e., only slightly above 3$\sigma$) and was not detected in the
TM1 configuration alone (detection only when combining TM1+TM2). The CO(3-2) line contours are from 3$\sigma$ to
  192$\sigma$, and follow each other by factor 2 multiplication.
}
\label{fig:1365-1433}
\end{figure*}

\section{Results}
\label{res}

\begin{figure*}[h!]
\centerline{
\begin{tabular}{c}
  \includegraphics[angle=-90,width=17.5cm]{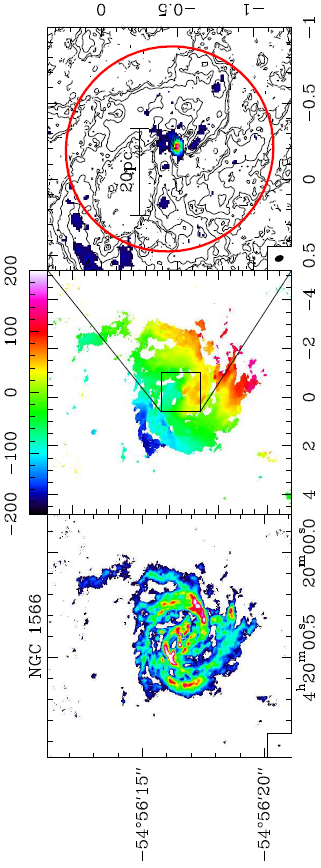}\\
\includegraphics[angle=-90,width=17.5cm]{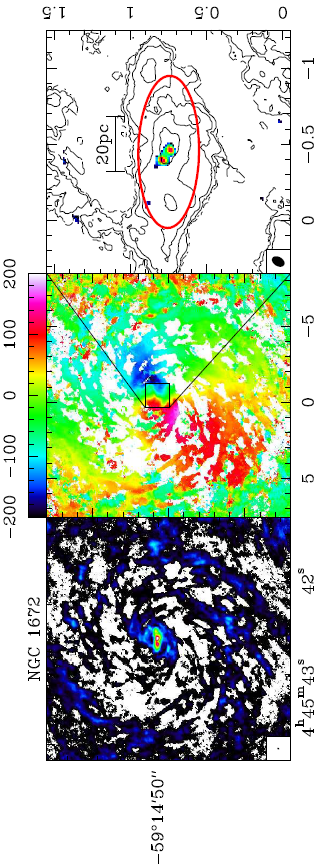}\\
\end{tabular}
}
\caption{{\it Top}: Same as Fig. \ref{fig:613-1326}, but  for NGC~1566
  (in this case  TM2+TM1 instead of TE+TM1 in the  first two panels). 
  The contours are from 5$\sigma$ to
  160$\sigma$, and follow each other by factor 2 multiplication.    {\it Bottom:}
  Same, but for NGC~1672. The contours are from 3$\sigma$ to
  192$\sigma$, and follow each other by factor 2 multiplication.
}
\label{fig:1566-1672}
\end{figure*}

Figures \ref{fig:613-1326} to  \ref{fig:1808} display the CO(3-2) flux distribution (moment 0)
together with the velocity field (moment 1) of the seven targets; they demonstrate the wide variety
of morphologies encountered. There are both resonant rings and spiral arms, and 
six galaxies host a molecular torus, either nearly face-on or more edge-on. 
 We define a torus by the smallest decoupled circumnuclear structure in the CO(3-2) line, and 
 from its morphology and kinematics. 
The morphology decoupling means that a nuclear disk or ring is clearly detached
from the outside spiral structure or outer rings, and the corresponding kinematic
displays a strong velocity gradient compared to the outside structure.
In some cases the kinematic major axis of the inner torus is not aligned
with the kinematic major axis of the outside structure (e.g., NGC~613, NGC~1566, NGC~1672 and
NGC~1808).  The HCO$^+$(4-3) line, when clearly detected, supports
 this definition (see Fig. \ref{fig:613-HCO}).
 The dust continuum may also trace this circumnuclear structure,
 but it is weaker and tends to be less extended radially. Also, the central point source, coming
 in general from synchrotron AGN emission,  perturbs its morphology. We emphasize that
 the radial extent of the torus may depend on the tracer considered, as already found in NGC~1068
 \citep{Garcia-Burillo2016, Garcia-Burillo2018}.
Most of the time there
are departures from symmetry, and the central torus is slightly offset from the barycenter
of the 1~kpc-structure. In two cases (NGC~613 and NGC~1566) there is a clear trailing spiral
inside the inner Lindblad resonance (ILR) ring, a ``smoking gun'' signature for ongoing fueling of the central black hole.
The main morphological and kinematical features are described for each galaxy below.

\subsection{Continuum emission and AGN position}
\label{continuum}

All galaxies were observed with their phase center coinciding with the
position given in NED. However, we now have  a high spatial resolution, and
when the nucleus has a strong continuum point source, it is possible to refine the position
of the central black hole, expected to be the source of strong radio emission.
A central continuum point source has been detected in all galaxies, except in NGC~1433
where we had a tentative detection in our previous work \citep{Combes2013},
but were not able  to subsequently confirm it. We adopt, however, the same previously determined position
for the center, which corresponds to the center of the inner nuclear ring.
For all others, the center position adopted is the pixel of maximum continuum emission,
determined with an error equal to the TM1 beam indicated in Table \ref{tab:config},
typically 0.1\arcsec.
These positions are displayed in Table \ref{tab:sample}, and  in the following
are considered to coincide with the actual AGN nucleus.
These positions are now better determined than the optical nuclei defined by the
maximum brightness in the HST images of Fig. \ref{fig:HST}, which  have an astrometry
accurate to within $\le$1\arcsec. Our adopted positions also coincide most of the
time with the maximum  velocity dispersion in the CO(3-2) line (except for NGC~1365
and NGC~1566).

\begin{figure*}[h!]
\centerline{
  \includegraphics[angle=-90,width=17.5cm]{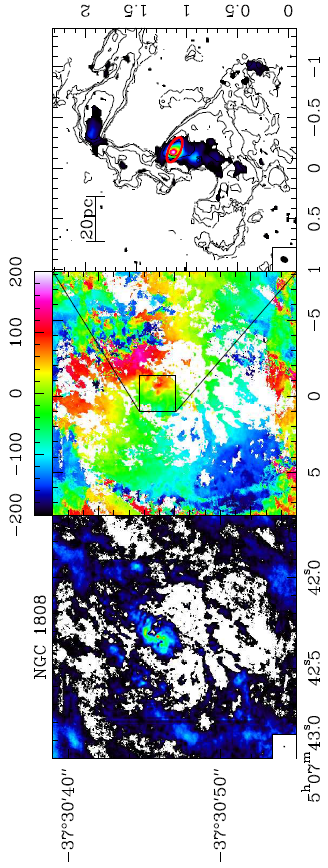}
}
\caption{Same as Fig. \ref{fig:613-1326}, but for NGC~1808. The contours are from 3$\sigma$ to
  384$\sigma$, and follow each other by factor 2 multiplication. The color scale in the middle panel is in km/s.
}
\label{fig:1808}
\end{figure*}

There is also the possibility that some continuum emission from the unresolved
central source is coming from the inner torus around the AGN.
Most galaxies in the sample are low-luminosity AGN  in which the dust emission close to the
nucleus is expected to emerge between 10 and 300$\mu$m \citep{Casey2012}. We expect that the continuum point source
at 0.8mm is dominated by synchrotron emission  \citep{Prieto2010}.
Around the unresolved continuum sources, figures \ref{fig:613-1326} to \ref{fig:1808} show that
we detect extended emission which might be dominated by dust emission
from the inner tori; however, these tori are  better traced by the CO line emission.

\subsection{Molecular gas distribution and morphology}
\label{morpho}

{\bf NGC~613} shows an incomplete ring with high-density contrast, as seen in our large-scale map of 
CO(3-2) (Fig. \ref{fig:613-1326}), and emission from a clear trailing two-arm spiral structure in the 
circumnuclear disk (CND).
This galaxy has a typical ILR 
nuclear ring, of radius 3.5 arcsec (or 300~pc), just inside the two characteristic leading dust 
lanes of the bar, which are tangent to the ring. Inside the ring, there is a central molecular component, 
or CND of radius $\sim$ 1\arcsec = 83~pc. 
Water masers have been detected in the nucleus \citep{Kondratko2006}.
\cite{Miyamoto2017} have mapped the ring in CO(1-0) and CO(3-2) with ALMA at 0\farcs7 and 0\farcs4, 
respectively, and found a clumpy ring, globally regular, but with spots of active and efficient star formation. 
ALMA finds a continuum jet at 95~GHz with a PA=20$^\circ$ \citep{Miyamoto2017}, which corresponds 
to the 4.9~GHz and 14.9~GHz jets \citep{Hummel1992}, close to the minor axis of the ring. 
At 350~GHz, the central continuum source remains unresolved at our resolution of 0\farcs09. The 
negative slope found for the flux over frequency $\alpha\sim$ 0.6 is compatible with synchrotron 
emission, with a small fraction of free-free.  

The ring reveals two breaks into two winding spiral arms to the  north and south. 
The bar has a position angle PA=130$^\circ$ and is delineated on its leading sides
by two dust lanes: one  runs from the SW of the nucleus to the east side of the bar (PA=130$^\circ$),
and the other from the NE of the nucleus to the west side of the bar (PA=-50$^\circ$)
(see Fig.\ref{fig:HST}).
The spiral arms are the beginning of these characteristic dust lanes along the bar: 
they are the contact points between the tangent lanes and the ring.

 Dense gas ($> 10^6$ cm$^{-3}$) is detected in various lines of HCN, HCO+, CS, and SiO at 
the edges of the jets \citep{Miyamoto2017}. We find that at high resolution with HCO$^+$(4-3)
(Fig. \ref{fig:613-HCO})
the dense molecular gas is very concentrated towards the nucleus, with skewed kinematics, 
suggesting an outflow along the jets; this will be described in a future paper \citep{Audibert2018}. 
 A first look at it can be seen in Fig. \ref{fig:mod-613}, where both the velocity
field and velocity dispersion are perturbed.
The outflow has already been suggested by the high-velocity dispersion of the 
[FeII] line along the radio jet \citep{Falcon2014, Davies2017}.

A coherent chronology of star formation possibly driven by the gas motions in the ring was sought by
\cite{Boker2008} with SINFONI images in Br$\gamma$, HeI, and [FeII] 
(spectral resolution R=2000, and without adaptive optics). They identified clumps along the ring,
 color-coded according  to the different star formation phases. These authors
searched for a coherent chronology of star formation, starting from the dust lanes, fueling gas 
into the ring. At least in the southern part of the ring the expected sequence of star formation 
was indeed observed: the hottest stars were found near the contact point
(defined above between the dust lane and the ring), and then fewer hot stars were found 
along the ring. The star formation scenario is not the random one, where new stars light up like 
a popcorn model. If gas is inflowing from the bar dust lanes into the ring, as expected from gravity 
torques \citep{Garcia-Burillo2005}, there must also be  an inflow in the CND due to the nuclear 
trailing spiral, as already observed and interpreted for NGC~1566 \citep{Combes2014}. 
The detailed nature of the gas flows in NGC~613 will be studied in a future work.
Inside the nuclear spiral structure, there is a very dense and compact (radius $\sim$ 14~pc) 
rotating component, which might be interpreted as the molecular torus. 
The torus is also distinguished by a kinematic decoupling, with a major axis PA=0$^\circ$,
(better seen in HCO$^+$(4-3)), while the outside axis is PA=105$^\circ$.
The excitation of the ionized gas in the torus is dominated by shocks \citep{Davies2017}.

{\bf NGC~1326} is a lenticular barred galaxy with a contrasted ring at ILR, of radius 5.7\arcsec = 410 pc. 
\cite{Garcia-Barreto1991} detected radio continuum in the ring at 20, 6, and 2cm 
with the VLA at $\sim$ 3'' resolution, and also a molecular mass of 2.7 10$^8$ M$_\odot$ from CO emission
with the SEST-15m telescope.
There is no evidence of a strong central continuum point source at centimeter wavelengths. 
At 350 GHz, we detect a weak continuum source, coinciding with the maximum of the CO(3-2) emission 
(Fig. \ref{fig:613-1326}). About 83\% of the H$\alpha$ emission of the galaxy is coming from this 
nuclear ring \citep{Crocker1996}. There are no conspicuous dust lanes along the bar, and the 
ring is not interrupted by spiral structure at their contact. The ring in CO and also in H$\alpha$ 
is oriented at PA=90$^\circ$ roughly perpendicular to the bar (PA=20$^\circ$), suggesting either 
that the gas is on x2 orbits, or that 
it forms a decoupled nuclear ring, with a pattern speed higher than that of the primary bar. 
In any case, this means that the primary bar pattern speed is slow enough to allow the presence of 
two ILRs.
When the rotation curve of the galaxy  \citep{Garcia-Barreto1991, Storchi1996} is considered, 
the ring corresponds to the inner ILR. 

The HST image in H$\alpha$ \citep{Buta2000} reveals a marked asymmetry of the ring, 
which we do not see in CO(3-2) with ALMA. While the 14''-diameter (1kpc) ring  overlaps
very well in H$\alpha$ and CO(3-2) emissions, it is depleted on its west side only in H$\alpha$.
This must therefore be due to strong dust extinction in the  west side. The ring is made of 
hundreds compact sources, which could be star clusters, and also some diffuse emission, 
tightly wrapped and spiral in character. This spiral structure is  seen in our CO(3-2) high-resolution map as well. 
Stellar populations were studied from U,B,V, I colors and H$\alpha$-filter images by \cite{Buta2000}.
The age range of the stellar population in the ring is 10-200 Myr and the derived star formation 
rate is 1 M$_\odot$/yr \citep{Buta2000}. Inside the nuclear ring, the CO emission reveals a central 
CND, of radius $\sim$ 0.3\arcsec = 21~pc. This central structure appears slightly more inclined 
on the sky than
the main galaxy, and is likely  a tilted torus.

{\bf NGC~1365} is a strongly barred spiral galaxy with the characteristic dust lanes delineating the bar,
which are particularly contrasted and curved. As an archetypal barred galaxy, it has been extensively
observed \citep[e.g.,][]{Lindblad1999}. While the Seyfert 1.8 nucleus is obvious in optical and X-rays 
\citep{Nardini2015}, it is hardly seen in radio. \cite{Stevens1999} discuss a marginal radio jet, the 
center of which corresponds to the central X-ray position. 

The main part of the radio emission at centimeter wavelengths  is the nuclear ring, with hot spots 
corresponding to star formation. 
\cite{Sakamoto2007} have mapped in CO(2-1) and isotopes a mosaic with SMA at 2\arcsec resolution,
that shows clearly a 2~kpc-extent oval ring, connected to the leading dust lanes, characteristics of the bar morphology.
The AGN has only a small contribution ($\sim$5\%) to the central infrared emission, which is
dominated by star formation \citep{Alonso2012}.

At 350 GHz, we detect a central continuum point source, as can be seen in Figure \ref{fig:1365-1433}. 
Our large-scale CO(3-2) map reveals the contrasted nuclear ring of radius $\sim$ 9\arcsec = 770~pc.  
Inside this ring, which corresponds to the ILR of the bar, we detect a more compact molecular 
component, a CND with a ring shape, of radius 0\farcs3 = 26~pc. This rotating ring just encircles 
the central continuum source, and might be interpreted as the molecular torus.  

\cite{Lena2016} have recently presented optical integral field spectroscopy for the inner 
6\arcsec. They find evidence for a fan-shaped blueshifted outflow in [NII] and H$\alpha$ 
kinematics, corresponding to the outflow in a cone seen with the [OIII] emission lines, 
extending at more than 1~kpc from the center along the
minor axis \citep{Hjelm1996, Venturi2017}.  From gravity torques, it was possible to show that 
the gas is inflowing to the center, driven by the bar, on a timescale  of 300 Myr 
\citep{Tabatabaei2013}. We will discuss in a future paper the possibility of an outflow in the 
molecular component \citep{Audibert2018}.

{\bf NGC~1433} is a strongly barred spiral galaxy, nicknamed the “Lord of Rings” because of the 
presence of clear nuclear, inner, and outer rings at the bar resonances \citep{Buta1996}. 
Our previous cycle 0 ALMA observations in Band 7, with 0\farcs5 resolution, have revealed that 
inside the nuclear ring at $\sim$ 400~pc, there is also an inner nuclear ring at $\sim$ 200~pc, 
corresponding to the inner ILR \citep{Combes2013}. These observations revealed also a mild molecular 
outflow along the minor axis, of 7 M$_\odot$/yr, the smallest molecular outflow ever observed in the 
Local Universe. It is possible that the outflow in a previous more powerful
phase has destroyed a potential torus. The tentative continuum point source near the center 
is not confirmed, and there is no evidence of any molecular torus.  

The only small velocity gradient in the center corresponds to the outflow
along the minor axis. This is supported  by the coincidence of the optical and near-infrared  emission 
in the center, showing no extinction \citep{Smajic2014}.  The CO(3-2) emission has a very 
filamentary structure at small scale, as can be seen in Figure \ref{fig:1365-1433}. 
The nuclear ring is the site of a
starburst \citep{Sanchez-Blazquez2011}, and the gas is
transiently stalled there though gravity torques \citep{Smajic2014}.

{\bf NGC~1566} is an intermediate barred spiral galaxy, possessing 
nuclear, inner and outer rings at resonances \citep{Aguero2004}.
The broad lines detected in the nucleus, and the observed variability
are typical of a Seyfert 1 \citep{Alloin1985}. It appears that this low-luminosity AGN appears has been 
increasing its activity
over the last few hundred years; the line excitation is much higher in the BLR than along the NLR cone
\citep{Baribaud1992, Reunanen2002, Smajic2015}. 

Our previous cycle 0
ALMA observations have shown that the molecular gas followed a trailing 
spiral inside the nuclear ring, fueling the central black hole. 
The gravity torques due to the bar on the gas change sign
at each resonance, and also change sign as the winding of spirals change from trailing to leading
\citep{Buta1996}. When the spiral is leading inside the ring located at the inner Lindblad resonance,
the torques are positive and the gas inside the ring is driven back to the ring.
Instead,  when the spiral is trailing, the torques are negative, the gas  loses angular momentum,
and is driven towards the nucleus.
The sense of winding of the spiral must be due to the gravitational influence
of the black hole itself \citep{Combes2014, Smajic2015}. The image at high resolution
confirms the trailing spiral structure in the nuclear disk (figure \ref{fig:1566-1672}).
The nucleus is also the site of young star formation, and of a consequent velocity dispersion drop,
also called $\sigma$-drop
\citep{DaSilva2017, Emsellem2001}. 

Inside the inner spiral fueling the nucleus, there
is a ringed structure of 0.7\arcsec = 24~pc radius, kinematically decoupled,
which can be considered  a molecular torus.
As can be seen in Fig. \ref{fig:1566-1672}, the torus kinematic major axis is vertical (PA=0$^\circ$),
while the rest of the disk has a kinematic major axis PA=60$^\circ$.

{\bf NGC~1672} is a strongly barred Seyfert 2 galaxy, with a high star-forming activity
in its center. The AGN activity is therefore hard to find through line diagnostics
\citep{Storchi1996, Kewley2000}.
\cite{Brandt1996} with ROSAT find a diffuse X-ray nuclear source, rather soft, compatible
with a superbubble interpretation (i.e., thermal emission from star formation).
 The star formation rate implies enough supernovae 
over a 10$^7$ yr period to blow a superbubble, as computed by \cite{MacLow1988},
although the X-ray gas pressure and density appear too high \citep{Brandt1996}.
While finding  two additional X-ray sources at the
bar extremities, \cite{DeNaray2000} conclude that the Seyfert 2 activity must be obscured
by a Compton-thick nucleus with a column density of at least 10$^{24}$ cm$^{-2}$.
The X-ray emission at the center appears diffuse and dominated by a starburst nucleus.
\cite{Jenkins2011} with the high resolution of Chandra are the first to
find a hard X-ray emission associated with the nucleus in addition to a ring. This confirms
that the galaxy is actually a low-luminosity AGN, a Seyfert 2.

The circumnuclear ring, of 5\arcsec = 275~pc radius, is conspicuous in the radio emission at 3cm from
the VLA, and also in the Spitzer 8$\mu$m band, tracing essentially the PAH
dust \citep{Jenkins2011}.
\cite{Diaz1999} provide the H$\alpha$ velocity field in the central 2kpc: it has a strong velocity gradient,
with a mass of 9 $\times$ 10$^8$ M$_\odot$ inside 125~pc.
The ring is located at the inner ILR, and is also quite contrasted in the CO(3-2) emission
(figure \ref{fig:1566-1672}). Inside the ring, some thin filaments join towards a central concentration,
which looks like a torus of radius  0.5\arcsec =27~pc seen more inclined than the large-scale disk. 
A kinematic decoupling helps to define the torus, with a kinematic major axis of the torus of PA=90$^\circ$,
and an outside axis of PA=135$^\circ$.
The continuum emission peaks just at the center.

{\bf NGC~1808} is a  barred starburst containing a Seyfert 2 nucleus. Optically, it is possible
to see the dust lanes expelled perpendicular to the major axis.
There is evidence of large-scale (1~kpc) outflows, likely due to the starburst, and NGC~1808 has
been classified as a superwind galaxy, similar to M82 \citep{Dahlem1990, Dahlem1994}. The outflow is
also seen in HI \citep{Koribalski1993}. Surprisingly, the outflow is not observed in the molecular
component at the small scales sampled here \citep{Audibert2018}. The outflow is thus apparently generated
at a large distance  from the nucleus ($>$ 300~pc), favoring a starburst-driven flow
over an AGN-driven one. 

Star formation is very active through hot spots aligned on a
ring of radius 10\arcsec = 450~pc \citep{Koribalski1996}.
Inside this ring, also contrasted in CO \citep{Salak2016}, there is a circumnuclear disk
or ring of radius 200~pc. Within this circumnuclear disk it is possible to see a trailing spiral
in the CO (figure \ref{fig:1808}), which is even more contrasted in the dense gas tracers: HCO$^+$(4-3),
CS. The continuum is a point source inside the nuclear spiral, and
may also correspond to a torus of radius 0.13\arcsec  = 6~pc.
The previous ALMA observations by \cite{Salak2017} have shown
the nuclear ring, and from the two contact points of the 
characteristic dust-lanes, two starting spiral arms. With a beam of
2\farcs6 = 100~pc, \cite{Salak2017} had insufficient spatial resolution
to see the trailing spiral inside the nuclear ring. They call
the nuclear ring a torus even if this component does not obscure the AGN.
In the following, we will define the torus as the smaller compact structure
of radius 6~pc inside the nuclear spiral. 
The kinematic major axis of the torus is PA=100$^\circ$, misaligned
with the outside axis of PA=140$^\circ$.

 \cite{Busch2017} have published a near-infrared IFU spectroscopy of the inner
600~pc with SINFONI in seeing-limited mode and R=1500-4000 spectral resolution.
 They determine an indicative black hole mass of 10$^7$ M$_\odot$.
The age of the stars on the ring is homogeneous and younger than 10~Myr. 
They find shocked H$_2$ warm gas near the nucleus, with  non-circular motions.
Although there is much gas streaming inside a radius of 1'' = 45~pc, there is
no strong sign of nuclear activity.

\begin{figure}[ht]
\centerline{
\includegraphics[angle=0,width=8.5cm]{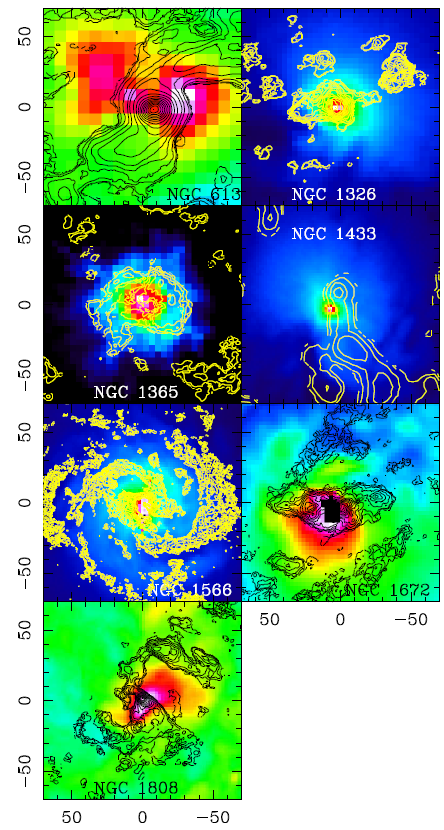}
}
\caption{ Red HST images (F814W) for the seven sample galaxies, zoomed into the 
central field of 70~pc radius, with the CO(3-2) contours overlaid. The axis labels are in pc.
}
\label{fig:HST-70pc}
\end{figure}

\subsection{Masses of possible tori}
\label{tori}

\begin{figure}[h!]
\centerline{
  \includegraphics[angle=-0,width=7.5cm]{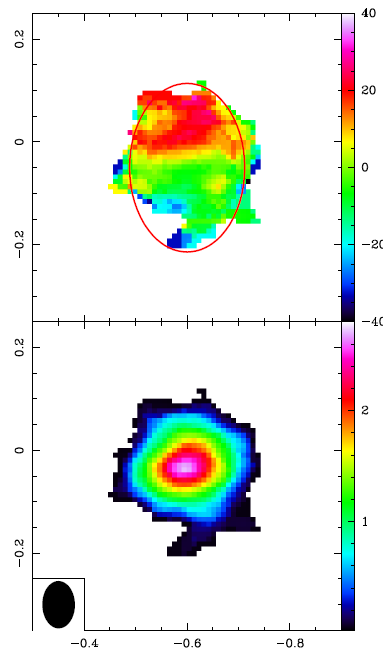}
}
        \caption{ First moments of the HCO$^+$(4-3) map of NGC~613.
        The color scale of the intensity at the bottom is in Jy/beam $\times$ km/s,
        and the velocity at the top is in km/s. The red ellipse at the top
        indicates the definition of the molecular torus. The RA-Dec scale is in arcsec.
}
\label{fig:613-HCO}
\end{figure}

In at least six  of the seven galaxies, there is evidence of a compact 
central ring or disk in CO emission, which
can be interpreted as a molecular torus.
 These structures can be seen in CO(3-2) contours overlaid
over HST images in Figure \ref{fig:HST-70pc}. Their kinematics 
are  also displayed  with models, starting from Figure \ref{fig:mod-1672}.  For NGC~1433, there is no evidence of a nuclear disk
rotating around the nucleus, although there is a piece of spiral arm
superposed on it.
The molecular nuclear disks or tori are clearly decoupled from the rest of the disks  in the kinematics as well.
This is seen in Figures \ref{fig:613-1326} to \ref{fig:1808}, and also particularly in the dense gas maps in
HCO$^+$(4-3) which will be discussed in future papers. The example of NGC~613 is 
displayed in Figure \ref{fig:613-HCO}.
In all galaxies, except for NGC~613 and NGC~1808, the size of the torus is much 
larger than the beam size (8 -- 26 times). For NGC~613 and NGC~1808, the torus
is only 4 times the beam, and the size reported in Table \ref{tab:torus}
has been deconvolved from the beam.

We  have identified in the CO(3-2) data cube these components through ellipse-fitting
with the adopted center as the continuum point source identified with the nucleus
(at the coordinates given in Table \ref{tab:sample}).
The sizes of the tori are illustrated by red ellipses in 
Figure \ref{fig:613-1326} and beyond.
From the integrated flux S(CO) dV
(Jy km/s) found within the region, the derived molecular mass is obtained through the formula
$$
{\rm L'}_{\rm CO} (K km/s/pc^2) = 3.25 x 10^7 \frac{S(CO) dV}{(1+z)}  \left(\frac{D_L}{\nu_{rest}}\right)^2, 
$$
where $\nu_{rest}$ =345.796 GHz, and D$_L$ is the luminosity distance in Mpc
\citep[e.g.,][]{Solomon2005}. The molecular mass, including helium, is then
derived from
$$
{\rm M}(H_2) = 4.36 {\rm L'}_{\rm CO}  {\rm R}_{13}
$$
assuming the standard CO-to-H$_2$ conversion factor of X$_{CO}$ = 2~10$^{20}$ cm$^{-2}$(K~km/s)$^{-1}$,
 applicable to Milky Way-like galaxies, and R$_{13}$ = L'$_{\rm CO1-0}$/L'$_{\rm CO3-2}$=2
 \citep[e.g.,][]{Tacconi2013}. Although this ratio was applied for high-z galaxies, it should
 also be appropriate to local galaxy nuclei, where the molecular gas is dense \citep[e.g.,][]{Braine1992, Dumke2001}. 
The actual excitation of the molecular gas in the tori, close to the AGN, is
yet not well known, and we refer to the above-mentioned papers, where our data on
HCN, HCO$^+$, and CS lines are discussed.

We display in Table \ref{tab:torus} the size and molecular
masses of the tori. They vary from 6 to 27~pc and 0.7 to 3.9 x 10$^7$ M$_\odot$.
In NGC~1326, 1672, and 1808, the torus is rather inclined on the sky plane,
and could obscure the nucleus. This is well correlated with the Sy 2 type.
The uncertainties on the torus size are determined from the extrema of the possible fits,
taking into account both moments zero and one,
and are never smaller than the beam size. The error bars on the integrated flux
and other properties follow from these definitions, assuming a fixed
 excitation ratio and  CO-to-H$_2$conversion ratio.

To gain insight on the possible H$_2$ column densities of the torus,
we  computed the average H$_2$ surface densities over the central beam
on the highest CO(3-2) resolution maps (TM1). The resulting H$_2$
colum densities are displayed in Table \ref{tab:torus}  together with the size 
of the beam
in parsec. These column densities are lower limits on the actual molecular
surface densities, both because of possible dilution in the beam,
and also possible missing flux. Conversely, they might be sometimes
an overestimation, if the CO(3-2) line becomes partially optically thin,
due to high gas temperature and large turbulent line-width.
 In either case, from these central column densities, we can derive the molecular mass
enclosed in the central beam, M$_{cent}$, displayed in Table \ref{tab:torus}.
For NGC~1365, the central beam is empty, and the column density
was taken from the peak surface density of the torus. The value of  M$_{cent}$ is in general
much lower than the expected central black hole mass, except in the case
of NGC~613. This does not affect our BH mass estimation, since 
the gas mass is taken into account for the gravitational potential.

It is interesting to compare the H$_2$ column densities inferred from our molecular 
measurements with the column densities
derived from X-ray absorption. While our derived column density
towards NGC~613 nucleus would suggest that it is 
 Compton thick, \cite{Castangia2013} find NH values from X-ray
spectral fitting that are lower, $\sim$4 $\times$ 10$^{23}$ cm$^{-2}$.
This may be explained by a clumpy torus. \cite{Nardini2015}
examine NGC~1365 during a Compton-thick state, but NGC~1365 is a well-known ``changing-look'' AGN
\citep{Matt2003}.
In NGC~1566, \cite{Kawamuro2013} find a NH column of
 $\sim$3 $\times$ 10$^{22}$ cm$^{-2}$ (with a 26\% covering fraction).
 For NGC~1672, \cite{Jenkins2011} find inconclusive evidence for a Compton-thick X-ray source,
although there is some evidence for NH $\sim$2 $\times$ 10$^{24}$ cm$^{-2}$ 
in agreement with our finding \citep{DeNaray2000}.
\cite{Jimenez-Bailon2005} find  NH $\sim$3 $\times$ 10$^{22}$ cm$^{-2}$
for the  hard nuclear source in NGC~1808.
In summary, our very high-resolution data are not directly comparable to the
X-ray measurements possible up to now. Our results are lower limits, allowing
quite high column densities in some regions of the torus, but the latter can
be quite clumpy.
 To be compatible with the X-ray estimations, the covering or surface filling factor 
has to be on the order of 0.01-0.1.

When there is a nuclear spiral structure, we have identified the torus 
as the nuclear component inside this structure (NGC~613,  1672, and 1808).
It is likely that the torus is replenished in gas through this spiral structure.
In that sense, the torus is the last axisymmetric structure before the nucleus.
It is however difficult to define a  typical torus in Seyfert galaxies: according
to the larger scale dynamics (resonances, nuclear rings, or spirals), the size of the torus
may vary widely.
Other tori in the literature  of even more varying sizes have been found:
if the torus in NGC~1068 is comparable (4~pc radius, \cite{Garcia-Burillo2016}, or
an extended one in the near-infrared of 27~pc radius \cite{Gratadour2015}), others are larger,
from 30-100~pc in the OH-megamasers of Mrk~231 \citep{Klockner2003}, or smaller
(1~pc for Circinus, \cite{Tristram2014}, as well as Seyfert with H$_2$O masers,
  with molecular transition at a fraction of a pc \citep{Madejski2000, Herrnstein2005}).
  It is not yet clear whether the obscuring structure must be a thick doughnut-shaped torus,
  or could be a warped thin disk \citep[e.g.,][]{Elitzur2006}. In NGC~3079, the torus has been
  identified with clumps elevated from the warped and flared disk, at 0.5~pc, forming a geometrically thick
  structure of  a few 10$^6$ M$_\odot$ \citep{Kondratko2005}.

\begin{center}
\begin{table*}
  \caption[]{Radii, masses, and inclinations of the molecular tori}
\label{tab:torus}
\begin{tabular}{lcccccccccc}
\hline
        Galaxy&Radius&S(CO)dV &M(H$_2$)$^{\mathrm{a}}$ &inc($^\circ$)&PA($^\circ$)&inc($^\circ$)$^{\mathrm{b}}$&Beam&logNH$_2$& M$_{cent}$& off-centering\\
& (pc)   &  Jy km/s & 10$^7$ M$_\odot$  & torus&torus& gal&    (pc) & (cm$^{-2}$)&  10$^6$ M$_\odot$& (pc) \\
\hline
NGC~613   &  14$\pm$3    &  56$\pm$20 & 3.9$\pm$1.4  & 46$\pm$7 &0$\pm$8   &36&6.2&25.3$\pm$.001&10.&42.\\
NGC~1326  &  21$\pm$5   &  18$\pm$2  & 0.95$\pm$0.1 & 60$\pm$5 &90$\pm$10 &53&5.3&23.9$\pm$.02&0.3&21.\\
NGC~1365  &  26$\pm$3   &  10$\pm$3  & 0.74$\pm$0.2 & 27$\pm$10&70$\pm$10 &63&6.3&23.5$\pm$.01&0.&86.\\
NGC~1433  &  --         & --         & --           & --       & --       &67&2.9&23.5$\pm$0.1&0.04& --\\
NGC~1566  &  24$\pm$5   &  72$\pm$10 & 0.88$\pm$0.1 & 12$\pm$12&30$\pm$10 &48&1.7&24.5$\pm$.01&0.1&7.\\
NGC~1672  &  27$\pm$7   &  80$\pm$9  & 2.5$\pm$0.3  & 66$\pm$5 &0$\pm$10  &28&4.0&24.3$\pm$.01&0.4&27.\\
NGC~1808  &  6$\pm$2    &  46$\pm$6  & 0.94$\pm$0.1 & 64$\pm$7 &65$\pm$8  &84&3.1&24.6$\pm$.004&0.5&58.\\
\hline
\end{tabular}
\begin{list}{}{}
\item[$^{\mathrm{a}}$] obtained with the
standard CO-to-H$_2$ conversion ratio, and
R$_{13}$ = L'$_{\rm CO1-0}$/L'$_{\rm CO3-2}$=2 
\item[$^{\mathrm{b}}$] from Hyperleda\\
 -- NH$_2$ for NGC~1365 is the peak of the torus, but the central beam is empty\\
 -- NH$_2$ for NGC~1433 is only that of the central arm\\
 -- The errors on NH$_2$ do not include the CO-to-H$_2$ conversion factor
\end{list}
\end{table*}
\end{center}

\subsection{Determination of the mass of the black hole}
\label{BHM}

With the obtained ALMA spatial resolution, up to 2~pc  for NGC~1566, we
are now reaching the sphere of influence (SoI) of the black holes in our nearby
Seyfert galaxies.  The SoI is defined as the region inside which
the gravity of the black hole of mass M$_{BH}$ dominates the gravitational potential of the host galaxy.
There are two possible expressions for the SoI, one is from \cite{Merritt2004},
when the stellar mass inside the SoI1 M$_*$(r $<$SoI1) is twice  M$_{BH}$, and the second 
involves the central velocity dispersion of the stars $\sigma_v$,
i.e.,  SoI2=GM$_{BH}$/$\sigma_v^2$.
 We display both in Table \ref{tab:soi}.
 In the following, we also consider for the gravitational potential
and the computation of the black hole mass; the example of NGC~1068 is presented here in comparison with the seven galaxies. Our previous high-resolution ALMA observations of its molecular torus have also reached
the BH SoI \citep{Garcia-Burillo2016, Garcia-Burillo2018}.

\subsubsection{Model of the gravitational potential}
\label{sec:galfit}

To compute the stellar distribution of stars in each of our galaxies, we use the S$^{4}$G 3.6~\mics
infrared images and their galfit decomposition in bulge and disk components
from \cite{Salo2015}. More precisely, there are in most cases three components, including an additional bar component. These components are determined by their Sersic index,
 their magnitude, their effective radius, and the axis ratio. In two galaxies only,
 there is also a point-source component corresponding to  7\% and 4\% of the total
 mass respectively for NGC~1365 and NGC~1566. In these cases, we represented this component
 by an exponential, with effective radius equal to the PSF of the  3.6~\mics observations,
 which has a FWHM of 2.1\arcsec \citep{Salo2015}. All disks are exponential. As for bars,
 the adopted distribution is the modified Ferrers ellipsoids \citep{Ferrers1877},
 with a surface density varying as [1-$\left(\frac{r}{R_{bar}}\right)^2$]$^2$, where R$_{bar}$
 is its outer truncation radius.  We adopt a mass-to-light ratio of
 M/L = 0.5 M$_\odot$ /L$_\odot$ for this  3.6~\mics band \citep[e.g.,][]{Sani2011, McGaugh2014, Lelli2016}.  
 All the adopted mass
 components for our seven galaxies, with NGC~1068 added for  comparison, are displayed in Table 
\ref{tab:s4g}.

%
\begin{center}
\begin{table}
        \caption[]{Stellar mass components from S$^{4}$G 3.6~\mics \citep{Salo2015}}
\label{tab:s4g}
\begin{tabular}{lccccc}
\hline
        Galaxy & Comp. & R$^*$  & M$_*$ &  Sersic n & B/T\\
       &     & (kpc) & (10$^{10}$ M$_\odot$)& & M$_*$\\
\hline
N~613  &  Bulge &  0.47 & 0.59 & 0.799&0.13\\
       &  Disk  &  3.8  & 3.4  & 1.0  &4.55\\
       &  Bar   &  5.9  & 0.56 &      &\\
N1326  &  Bulge &  0.43 & 0.63 & 1.167&0.32\\
       &  Disk  &  2.7  & 0.99 & 1.0  &1.98\\
       &  Bar   &  4.1  & 0.36 &      &\\
N1365  &  Bulge &  1.1  & 2.3  & 0.857&0.25\\
       &  Disk  &  8.2  & 5.0  & 1.0  &8.66\\
       &  Bar   &  7.8  & 1.3  &      &\\
       &  Nucleus& 0.25 & 0.06 &      &\\
N1433  &  Bulge &  0.31 & 0.20 & 1.379&0.14\\
       &  Disk  &  3.0  & 1.1  & 1.0  &1.46\\
       &  Bar   &  3.7  & 0.16 &      &\\
N1566  &  Disk1 &  1.0  & 0.36 & 1.0  &0.37\\
       &  Disk2 &  4.6  & 0.27 & 1.0  &0.97\\
       &  Bar   &  6.9  & 0.30 &      &\\
       &  Nucleus& 0.1  & 0.04 &      &\\
N1672  &  Bulge &  0.36 & 0.54 & 0.749&0.23\\
       &  Disk  &  3.3  & 1.5  & 1.0  &2.32\\
       &  Bar   &  4.0  & 0.28 &      &\\
N1808  &  Bulge &  0.36 & 0.81 & 1.029&0.39\\
       &  Disk  &  3.3  & 0.54 & 1.0  &2.09\\
       &  Bar   &  6.8  & 0.74 &      &\\
N1068  &  Bulge &  0.58 & 2.4  & 1.181&0.38\\
       &  Disk1 &  0.94 & 2.3  & 1.0  &6.30\\
       &  Disk2 &  7.0  & 1.6  &      &\\
\hline
\end{tabular}
\begin{list}{}{}
\item[$^*$] This is the effective radius R$_e$ for a bulge,
  the scale-length h$_r$ for an exponential disk, and
  the end radius R$_{bar}$ for a bar.
\item  M$_*$ is the sum of all stellar components 
        (bulge, disk, and bar) in 10$^{10}$ M$_\odot$
\end{list}
\end{table}
\end{center}

\begin{figure*}[h!]
\centerline{
  \includegraphics[angle=0,width=15cm]{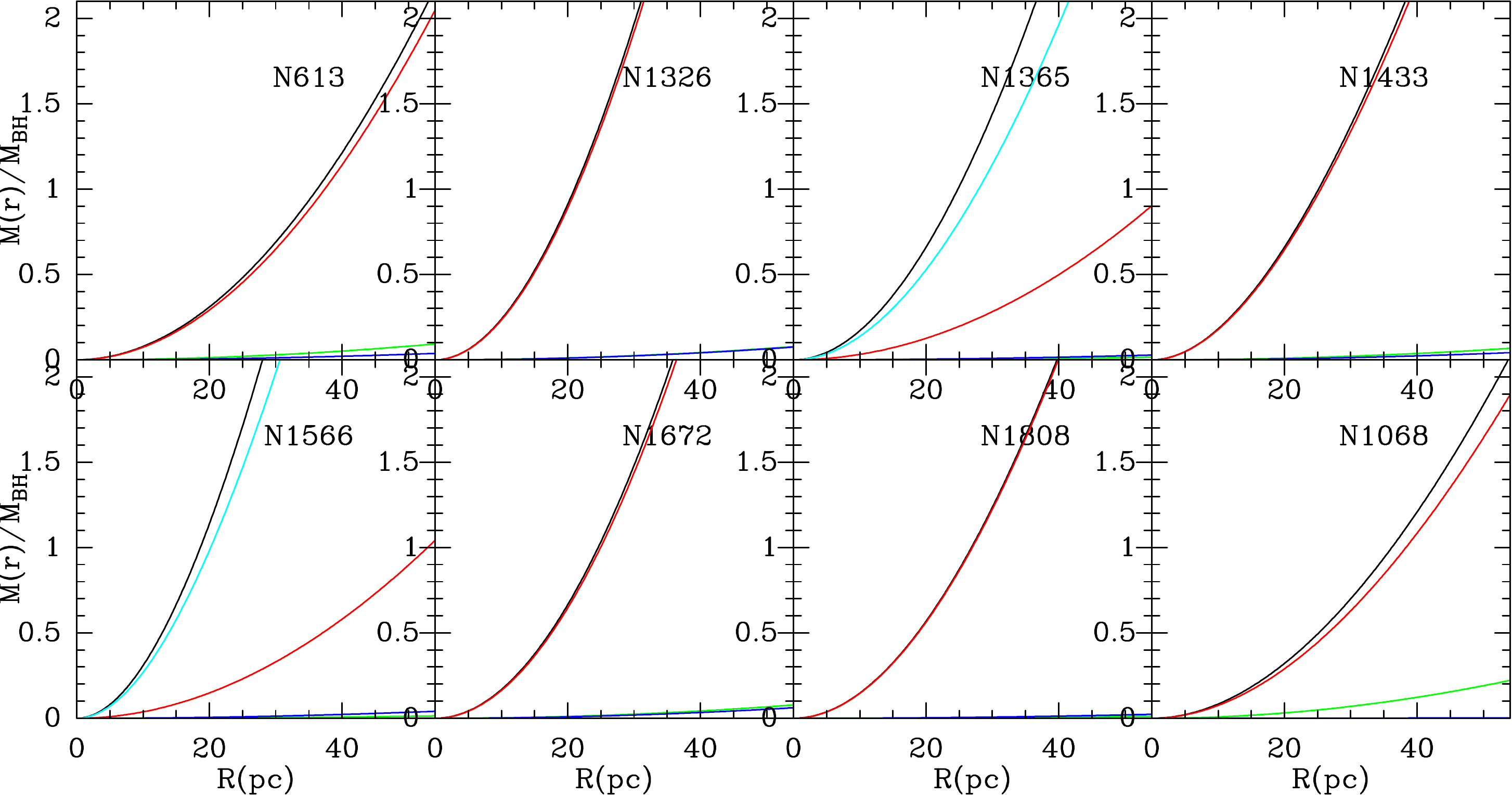}
}
\caption{Representation of the enclosed mass as a function of radius,
taking into account every component of the galfit 3.6~\mics model 
(Table \ref{tab:s4g}), with
their respective geometry (spherical bulges, and flat disks and bars)
for the eight galaxies. The vertical scale is normalized to their black hole
mass, estimated from their central velocity dispersion (Table \ref{tab:soi}).
The total mass is in black, the bulge in red, the disk in green, the bar in blue,
and a possible nucleus in cyan.
}
\label{fig:gal-mass}
\end{figure*}

\begin{center}
\begin{table}
  \caption[]{Central velocity dispersions and derived BH masses}
\label{tab:soi}
\begin{tabular}{lcccccc}
\hline
Galaxy&$\sigma_v$&Ref&  M$_{BH}$ &BH/bul &SoI1& SoI2\\
        & (km/s)   &   & log M$_\odot$  &  \%  &  pc &  pc\\
 (1)    &   (2)    &(3)& (4)            &  (5) & (6) & (7) \\
\hline
N 613  & 122$\pm$18   & (1)  & 7.57$\pm$.27  & 0.63 & 50 &11.\\
N1326  & 111$\pm$14   & (1)  & 7.40$\pm$.22  & 0.39 & 29 &8.8\\
N1365  & 141$\pm$19   & (1)  & 7.84$\pm$.26  & 0.30 & 33 &15.\\
N1433  & 113$\pm$3    & (2)  & 7.40$\pm$.03  & 1.25 & 36 &8.8\\
N1566  & 98$\pm$7     & (1)  & 7.13$\pm$.10  & 0.37 & 25 &6.2\\
N1672  & 111$\pm$3    & (2)  & 7.40$\pm$.03  & 0.46 & 33 &8.8\\
N1808  & 138$\pm$9    & (1)  & 7.79$\pm$.13  & 0.76 & 36 &14.\\
N1068  & 174$\pm$9    & (1)  & 8.23$\pm$.13  & 0.71 & 50 &25.\\
\hline
\end{tabular}
The references for the central velocity dispersion in the second column are:\\
(1) Hyperleda compilation \footnote{http://leda.univ-lyon1.fr/}\\
(2) \cite{Garcia-Rissmann2005} (cross-correlation method, CCM)\\
The SoI is the radius of the sphere of influence of the black hole,
the value in  Col. 6 is from Merritt's definition, M$_*$(r$<$SoI1) = 2 M$_{BH}$,
the value in Col. 7  is SoI2=GM$_{BH}$/$\sigma_v^2$.
\end{table}
\end{center}

To have a first estimation of the black hole masses, it is
interesting to exploit the well-known M$_{\rm BH}$-$\sigma$ relation for classical bulges
\citep[e.g.,][]{Tremaine2002,Marconi2003,Gultekin2009}.
We use the hyperleda compilation \citep{Makarov2014} for the
central velocity dispersion, and also \cite{Garcia-Rissmann2005}
when the data are not present in this data base.
To derive the BH mass, we use the recent M$_{\rm BH}$-$\sigma$ relation,
$$
log M_{BH}(M_\odot) = 8.5\pm0.05 + (4.41\pm0.29) log \left( \frac{\sigma_v}{200km/s}\right),
$$
compiled by \cite{Kormendy2013}.
The resulting masses are listed in Table \ref{tab:soi}, together with an indication
of the bulge-to-black hole mass ratio (according to the S$^{4}$G 3.6~\mics bulge decomposition)
and the derived estimation of the SoI radius. 
The derived mass models from Table \ref{tab:s4g}, including the BH masses of Table \ref{tab:soi}
are displayed in Figure \ref{fig:gal-mass}. The individual components from the galfit 
3.6~$\mu$m models are color-coded, and the vertical axis shows the SMBH versus
galaxy mass ratios. From these values, it is clear that the ALMA
observations can sample the dynamics inside the BH SoI in all galaxies.
It can be seen in Figure \ref{fig:gal-mass} that
the enclosed stellar mass is not dominant in the regions sampled by the molecular tori,
and the uncertainties on the determination of the stellar masses will
not have a large influence on our modeling.

Many studies have shown that pseudo-bulges and/or barred galaxies may be offset
from the main M$_{\rm BH}$-$\sigma$ relation \citep{Sani2011,Graham2011,Kormendy2013, Ho2014}.
Some of our galaxies may be in this category, 
and thus these first BH masses may be overestimated.
 It is therefore interesting to 
compare these values with several other BH mass estimations, such as the spiral pitch angle \citep{Davis2014},
or the Sersic index \citep{Mutlu2016}. These values are indicated in Table \ref{tab:bh2},
and are indeed about 1-30 times (on average 4 times) below the previous ones.

\begin{center}
\begin{table}
  \caption[]{BH masses from literature, and our derived ones}
\label{tab:bh2}
\begin{tabular}{lccccc}
\hline
Galaxy&  log M$_{BH}$ & log M$_{BH}$ &Ref& log M$_{BH}$ &log L$_{AGN}$\\
      & M$_\odot$     &  M$_\odot$ &     &  M$_\odot$ & erg/s   \\
 (1)  &        (2)    &    (3)     & (4) &     (5)    & (6)     \\
\hline
N~613  & 6.87$\pm$.27&  7.60$\pm$.35 & (1)  & 7.57$\pm$.15& 42.1\\
N1326  & 7.47$\pm$.17&  7.11$\pm$.33 & (2)  & 6.81$\pm$.2 & 40.7\\
N1365  & 6.05$\pm$.39&  6.30$\pm$.4  & (3)  & 6.60$\pm$.3 & 42.8\\
N1433  & 6.61$\pm$.37&  7.24$\pm$.4  & (4)  &  --         & 40.0\\
N1566  & 7.11$\pm$.32&  6.48$\pm$.2  & (5)  & 6.83$\pm$.3 & 41.4\\
N1672  & 7.08$\pm$.9 &  6.00$\pm$.6  & (6)  & 7.70$\pm$.1 & 39.3\\
N1808  & 6.74$\pm$.35&  7.20$\pm$.6  & (7)  & --          & 40.6\\
N1068  & 6.93$\pm$.37&  7.15$\pm$.1  & (8)  & 7.17$\pm$.2 & 44.7\\
\hline
\end{tabular}
\\
Column 2 lists BH masses estimated from
the spiral pitch angle \citep{Davis2014}, except for
NGC~1326, from the Sersic index \citep{Mutlu2016}.\\
Column 3 is from other estimations:\\
(1)  \cite{vandenBosch2016},  (2)  \cite{Mould2012}, (3) \cite{Risaliti2009},  (4) \cite{Smajic2014},
(5)  \cite{Smajic2015}, (6) \cite{Jenkins2011}, (7) \cite{Busch2017}, (8) \cite{Gallimore2001}\\
Column 5 displays our best fit estimates from the molecular gas dynamics. For NGC~1068,
the value comes from the H$_2$O masers  \citep[e.g.,][]{Lodato2003,Gallimore2016}.\\
Column 6 is the bolometric AGN luminosity (see Sect. \ref{sec:BHmass}).
\end{table}
\end{center}

From  our ALMA data on the six galaxies with well-defined kinematic tori
(the seven galaxies in  Table \ref{tab:torus}, but without NGC~1433), 
we can try to derive the CO line kinematics inside the SoI.
We use the position-velocity (PV) diagrams along the major axis,  as employed by the
WISDOM project \citep{Onishi2017, Davis2017, Davis2018} in deriving
the supermassive black hole masses in three early-type galaxies from the molecular
gas kinematics.
We postpone to future work the modeling of the full velocity field
of each galaxy because most objects are barred and subject to non-circular
motions (S-shape), as seen in Figs. \ref{fig:613-1326} to  \ref{fig:1808}.
As for the black-hole mass determination, we  only focus on the central part of the PV diagram
since the tori are frequently misaligned and tilted with respect to 
the main disk \citep{Fischer2013}.
To determine the PA of the possibly decoupled tori, we plotted PV diagrams for a
wide range of position angles (the first 36 separated by 10$^\circ$, then 
with a separation of 5$^\circ$ around the most probable PA), and selected the largest
velocity gradient. Only the cubes with the most extended ALMA configuration were
considered for this analysis (TM1).
The analysis of the PV diagram in NGC~1808 did not allow the construction of an unambiguous 
mass model;  the dynamics in the center is highly chaotic, 
probably due to the starburst and associated
feedback. We therefore analyzed only the five remaining galaxies
(the seven from  Table \ref{tab:torus}, but without NGC~1433 and NGC~1808),
 with the methodology described here.

\begin{figure*}[h!]
\centerline{
  \includegraphics[angle=0,width=15cm]{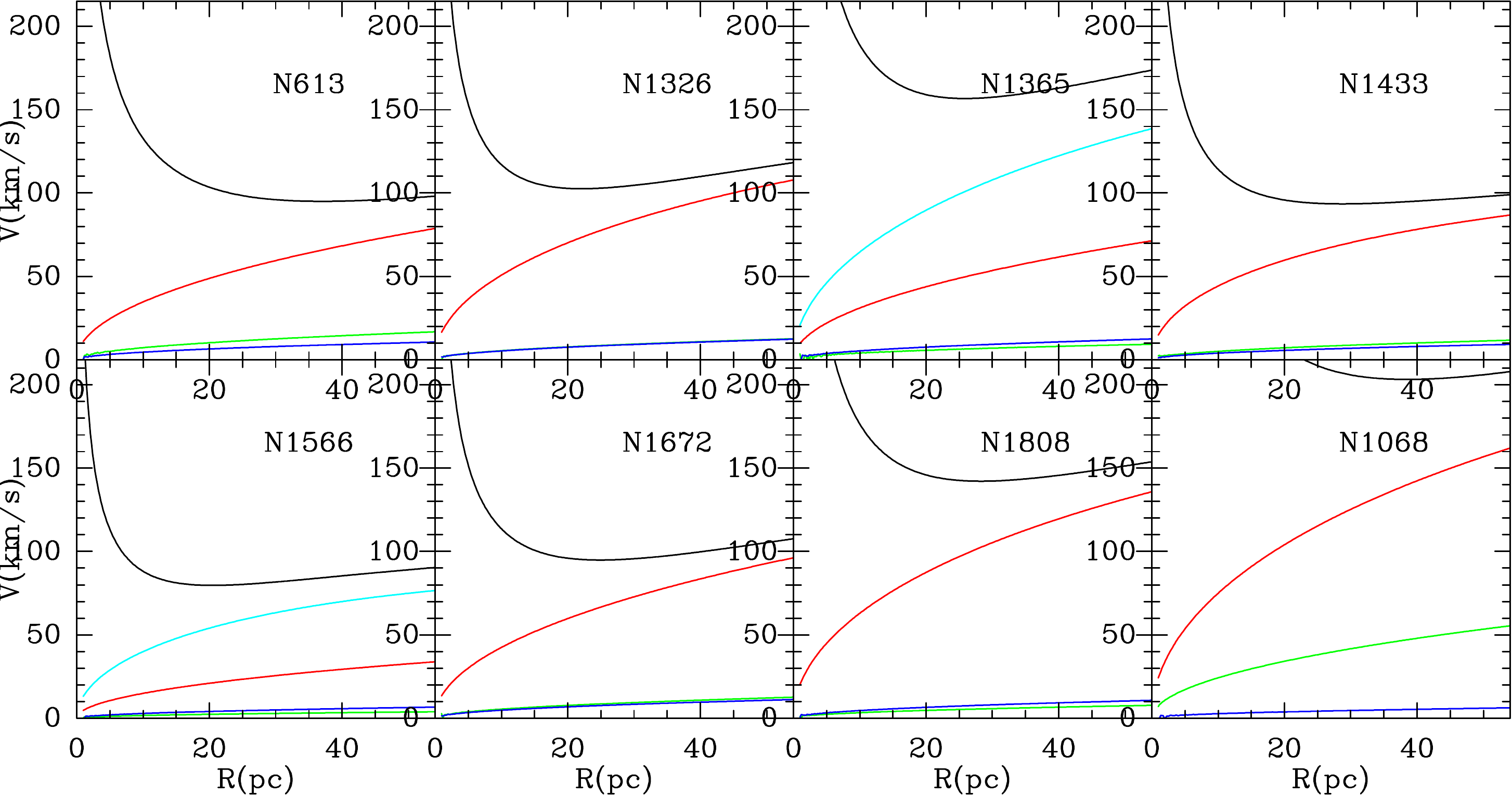}
}
\caption{Corresponding rotation curves for the eight galaxies,
derived from the mass models in Fig. \ref{fig:gal-mass}. The contribution
of the three or four components is color-coded as in 
         Fig. \ref{fig:gal-mass}. The contribution
of the black holes is included, with masses derived from the M$_{\rm BH}$-$\sigma$ relation
(Table \ref{tab:soi}).
}
\label{fig:gal-vit}
\end{figure*}

\subsubsection{Fitting the PV diagrams, and the moment maps}
\label{sec:PVfit}
To be able to predict the gas dynamics in the inner nuclear disk with more
flexibility, and with tilted decoupled dynamics,
we performed numerical simulations of gas particles in a selected potential.
We represent the gaseous disks by Miyamoto-Nagai disks of particles
\citep{Miyamoto-Nagai1975}, with
radial scales and heights corresponding to the observed tori.
We used typically one million particles to ensure sufficient statistics.
We plunge the gas disks into a potential made of the stellar distribution
determined by galfit to each galaxy, as described in the previous section.
The potential of the dark matter halo is negligible inside radii of 100~pc.
We take into account the gas mass to build the total gravitational potential.
The gas particles are distributed in circular orbits, in equilibrium with the total potential,
with a velocity distribution corresponding to a Toomre Q-parameter of 1. The ratio between
tangential and radial velocity dispersion was taken from the epicyclic theory
\citep[e.g.,][]{Toomre1964}.
It should be  noted that we simulate only axisymmetric disks. Here we consider  bar or spiral
perturbations to be secondary, and will study their effects in future papers.

Figure \ref{fig:gal-vit} illustrates the small-scale rotation curves predicted by our mass models where
the colors correspond to the same structural components as in Fig. \ref{fig:gal-mass}.
The quality of the potentials was checked by comparing at larger scales
the modeled rotation curve with previous observations. These rotation curves
come from H$\alpha$, HI, or low-resolution CO observations, and do not have  enough
spatial resolution to sample the curve below 100~pc, but give a calibration at 0.2-1~kpc.
Within these radii, the dark matter contribution is negligible.
The curves are generally good approximations to those on larger scales, with some scatter 
 which could come from the effect
of non-circular motions. As will be seen below, the influence of the large-scale
rotation has limited impact inside the sphere of influence of the black hole.

To compare the model to observations, we built data cubes by projecting the model
on the sky, with the best fit large-scale inclinations and position angles (see above), and 
computing the line-of-sight velocity distribution. We select the same pixel size as
the observed data cube (around 0.01 arsec according to each galaxy), 
and channels of 10 \kms, and the data were smoothed to the observed
beam (Table \ref{tab:sample}).  The sizes in pixels of the cubes were
between (180, 180, 60) and (360, 360, 60) to best
 sample the various tori studied.  Because the
 gas distribution is not homogeneous, but asymmetric and patchy, and this impacts
 the mass-weighted velocity at each observed beam, we normalized
 the model cube to the zeroth moment map of the CO observations,
 pixel by pixel in this 2D projection. This plays the role of  a multiplicative filter for
 our homogenous gas disks. This means that each CO spectrum at each position of the model
 is normalized to the observed integrated flux at this position.

\begin{figure*}[h!]
\centerline{
  \includegraphics[angle=0,width=16cm]{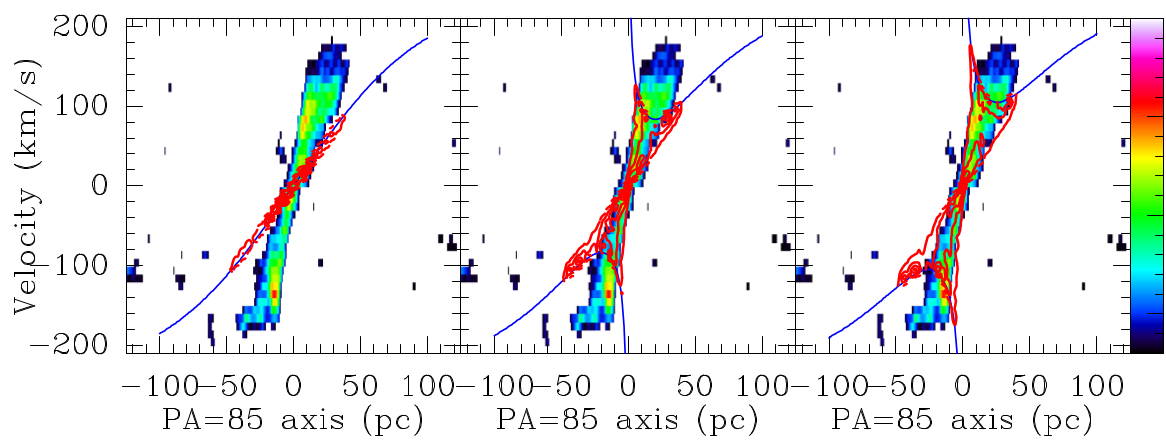}
}
\caption{Position-velocity diagram of the CO(3-2) line in NGC~1672, with a 
linear color scale (TM1 only).
Superposed in red are the contours of the model without any black hole
and torus inclination of 66$^\circ$ (left panel), with a black hole as derived from
the M$_{\rm BH}$-$\sigma$ relation (Table \ref{tab:soi})
of 2.5 x 10$^7$ M$_\odot$ with i= 66$^\circ$ (middle panel), and  the best fit: a black hole of 5.0 x 10$^7$ M$_\odot$, with i= 66$^\circ$ (right panel).
The mass model is that based on
the galfit decomposition, and the predicted circular velocity
is reproduced in blue lines (Fig \ref{fig:gal-vit}).
}
\label{fig:3pv-1672}
\end{figure*}

\begin{figure*}[h!]
\centerline{
  \includegraphics[angle=0,width=9.2cm]{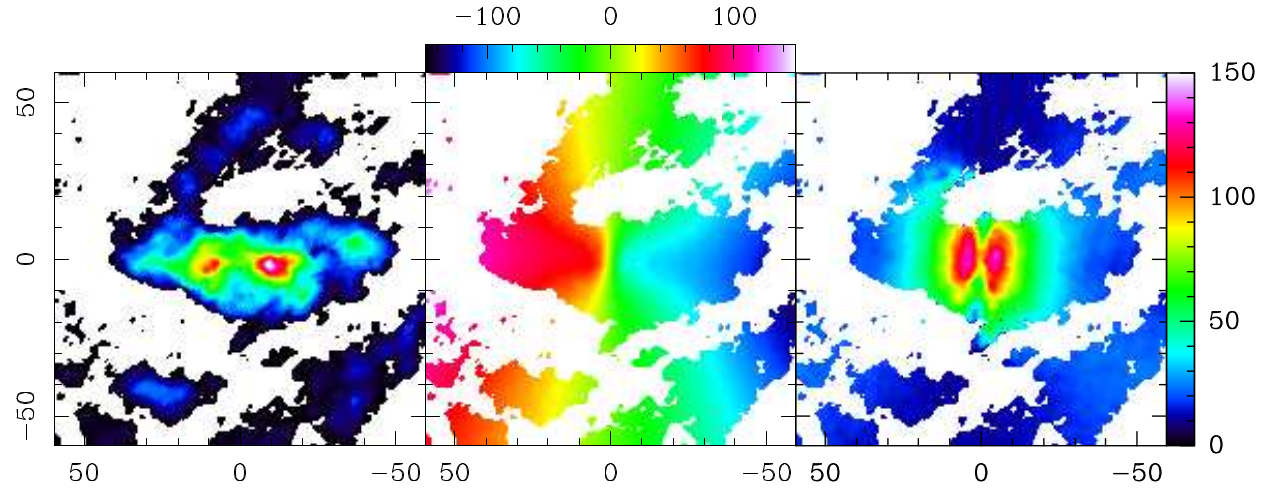}
    \includegraphics[angle=0,width=9.2cm]{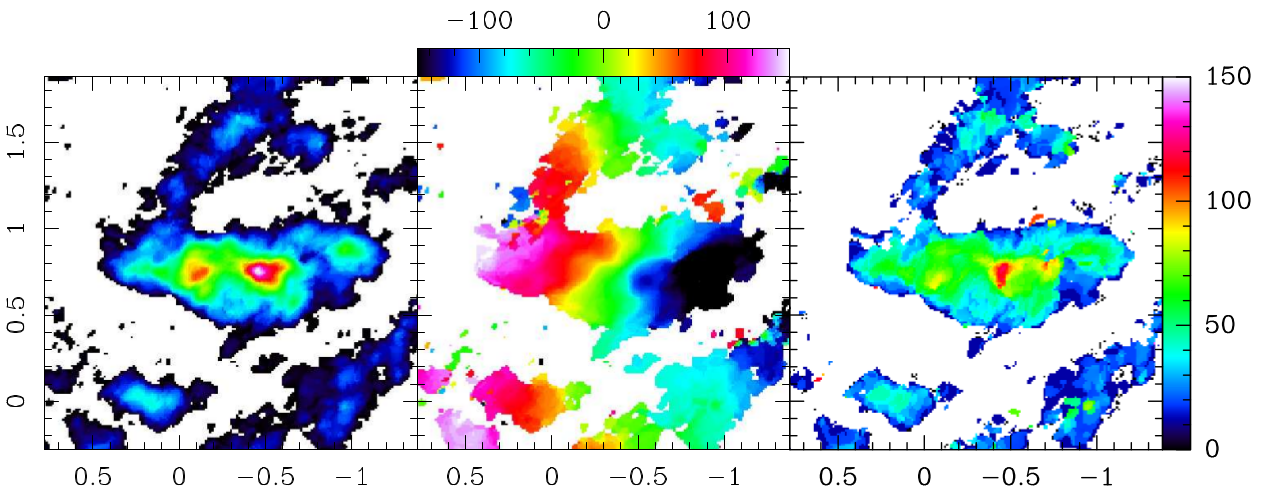}
}
\caption{{\bf Left}: Three first moments of the model cube for NGC~1672.
  The RA-Dec offsets are in parsec. The velocity color scales are  in km/s.
  The cube corresponds to the best PV fit, with a BH mass of  5.0 x 10$^7$ M$_\odot$,
  and an inclination of  66$^\circ$.
  {\bf Right}: Three moments for the observations (TM1 only).  The RA-Dec offsets are in arcsec
  with respect to the phase center.
}
\label{fig:mod-1672}
\end{figure*}

We tested the methodology on the most regular PV diagram obtained,
that of NGC~1672, as shown in Figure \ref{fig:3pv-1672}. The three
first moments of the model cube can be seen in Figure \ref{fig:mod-1672}.
The first step is to run, as a comparison, a model with no black-hole, which is
displayed in the left panel of Fig. \ref{fig:3pv-1672}. Then as a first estimate, the
value derived from a standard M$_{\rm BH}$-$\sigma$ relation, as given in Table
\ref{tab:soi}. We explore around this by small increments
in M$_{BH}$, which will allow us to obtain the best fit, maximizing the
overlap between the contours of the model and the observed map
in the PV diagram, and also fitting the three first moments.
Typically a dozen values are explored for each parameter,
which yield the best fit with error bars displayed in Table \ref{tab:bh2}.

\begin{figure}[h!]
\centerline{
  \includegraphics[angle=0,width=8cm]{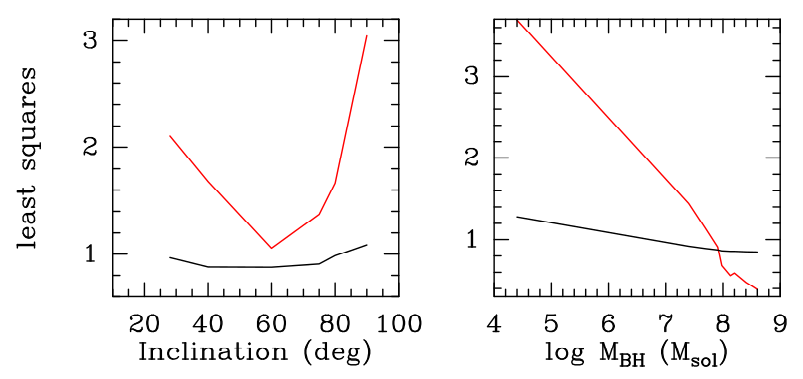}}
\caption{Least-squares fitting between the observed and modeled
  PV diagram (black curve) and 3D cube (red curve) as a function
  of inclination of the torus (left) and the black hole mass (right).
        The criterion is defined in Sect. \ref{sec:PVfit}.
  The curves have been slightly translated vertically for clarity.
  If there is a minimum for inclination, there is none for M$_{BH}$,
  and larger masses are preferred, as for NGC 1672.}
\label{fig:least}
\end{figure}

Two parameters were varied: the inclination of the central molecular disk/torus,
and the mass of the black hole.
To quantify the goodness of fit, we computed the least-squares values
summing the difference of all pixels between the observed and model maps,
either at 2D on the  PV diagram, or the moment maps, or at 3D on the cubes. We
concentrate on the nuclear part, with 3 10$^4$ pixels in 2D or 2 10$^6$ pixels in 3D,
corresponding to a region 120~pc in diameter, with a resolution of 0.66~pc and 10km/s.
The model and observed maps are normalized to the same total flux over this region,
and the squares difference is weighted by the observed flux in this pixel.
The criterion is then to minimize the quantity:
$ \sum\limits_{pix} w (F_{obs}-F_{mod})^2 / \sum\limits_{pix} w F_{obs}^2$, with the weighting 
function $w=F_{obs}>0$.
The result for NGC 1672 is illustrated in Fig. \ref{fig:least} for the PV diagram
and the total cube.

The fit for the PV diagram and also for the 3D-cube tend
to point towards large masses for the black hole. However,
these high masses are not realistic, since they created a 
central velocity dispersion that is too large, as can be seen in Fig. \ref{fig:mod-1672}.
The best fit must therefore be a compromise between the PV diagram and the
velocity dispersion map.

\begin{figure*}[h!]
\centerline{
  \includegraphics[angle=0,width=16cm]{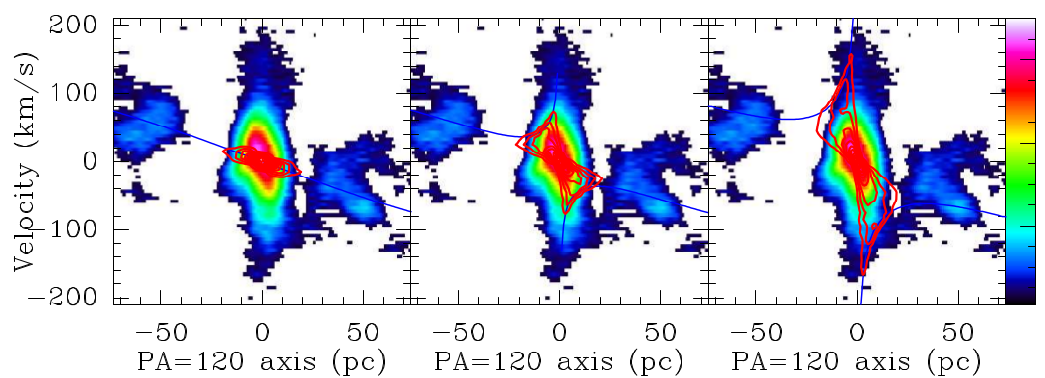}
}
\caption{Same as Fig \ref{fig:3pv-1672} for NGC~613,
without any black hole (left panel), with a black hole
of 7.4 x 10$^6$ M$_\odot$ (middle panel), and the best fit:
3.7 x 10$^7$ M$_\odot$, as derived from
the M$_{\rm BH}$-$\sigma$ relation (Table \ref{tab:soi}) (right panel).
The inclination of the torus is  i= 46$^\circ$.
The mass model is that based on
the galfit decomposition, and the predicted circular velocity
is reproduced in blue lines (Fig \ref{fig:gal-vit}).
}
\label{fig:3pv-613}
\end{figure*}

\begin{figure*}[h!]
\centerline{
  \includegraphics[angle=0,width=9.2cm]{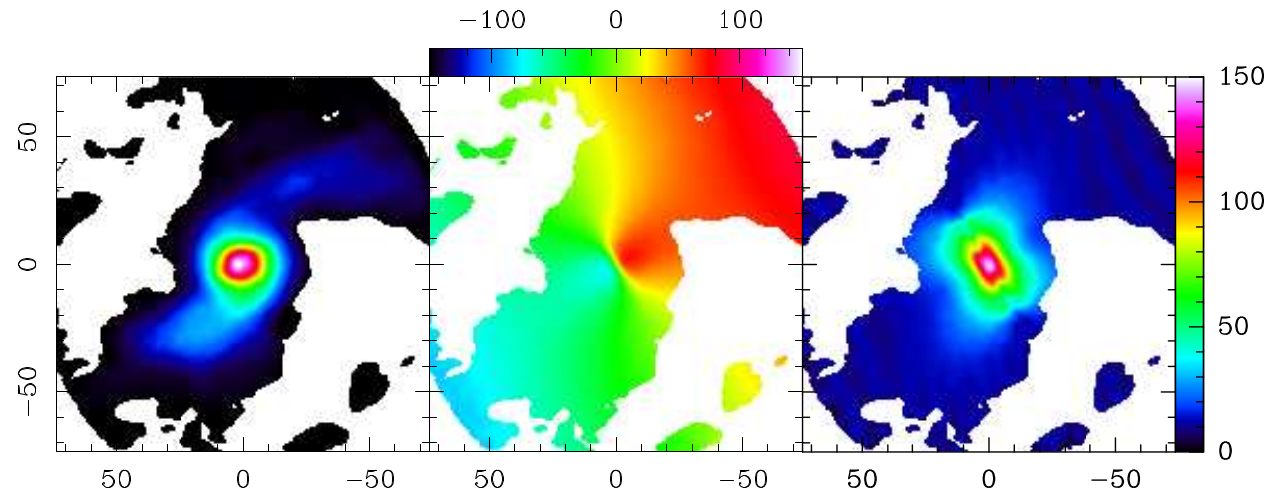}
    \includegraphics[angle=0,width=9.2cm]{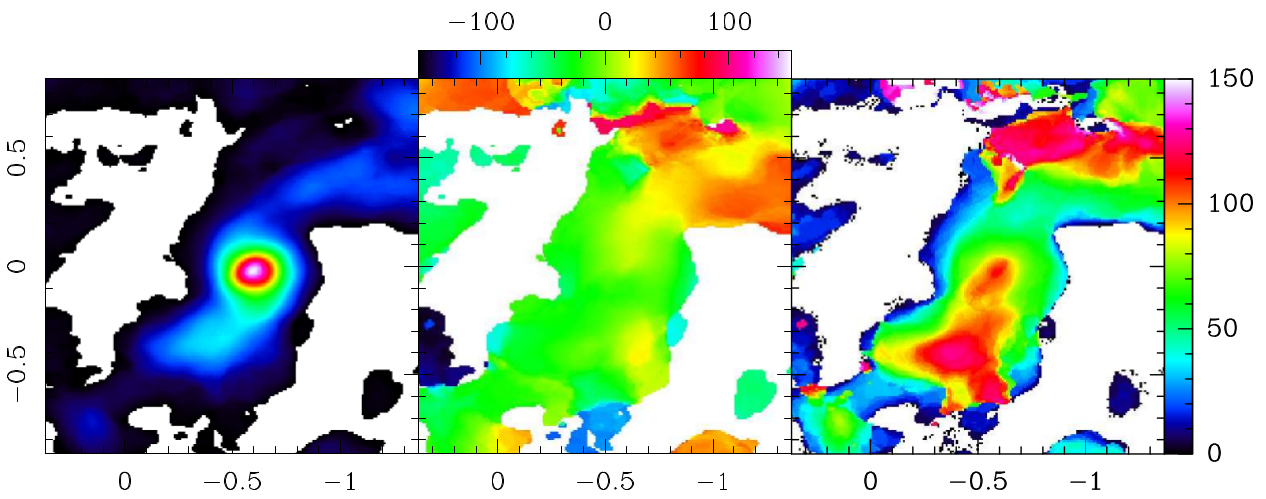}
}
\caption{ Same as Fig. \ref{fig:mod-1672} for NGC~613 (left is the model, right the observations).
  The cube corresponds to the best PV fit, with a BH mass of  3.7 x 10$^7$ M$_\odot$
  and an inclination of  46$^\circ$.
}
\label{fig:mod-613}

\end{figure*}

\begin{figure*}[h!]
\centerline{
  \includegraphics[angle=0,width=16cm]{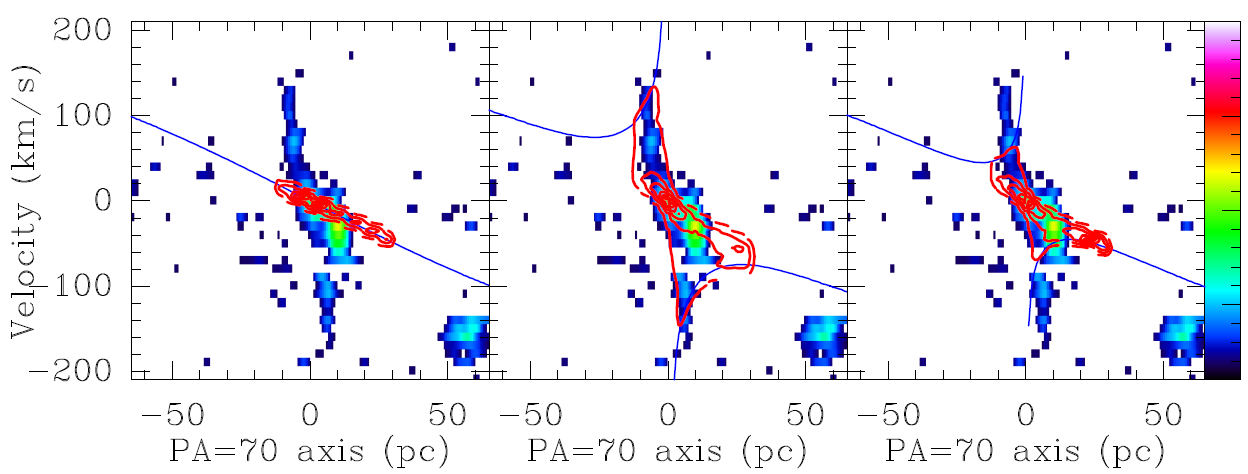}
}
\caption{Same as Fig, \ref{fig:3pv-1672} for NGC~1326,
without any black hole,  i= 60$^\circ$  (left panel); with a black hole
of 3 x 10$^7$ M$_\odot$  from  the M$_{\rm BH}$-$\sigma$ relation,  i= 60$^\circ$ (middle panel); and the best fit:
6.5 x 10$^6$ M$_\odot$, with  i= 60$^\circ$ (right panel). 
}
\label{fig:3pv-1326}
\end{figure*}

\begin{figure*}[h!]
\centerline{
  \includegraphics[angle=0,width=9.2cm]{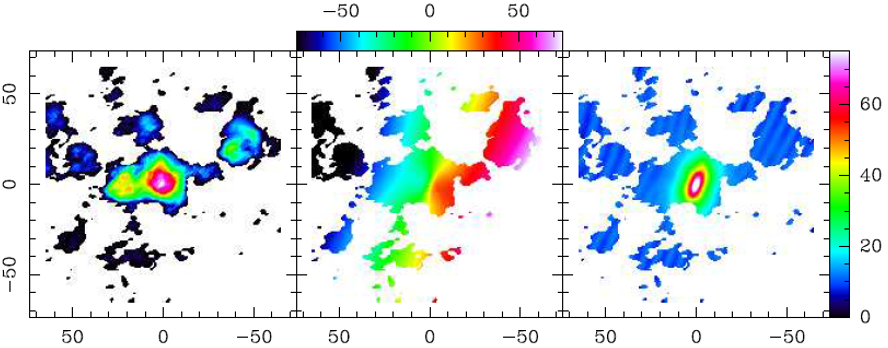}
    \includegraphics[angle=0,width=9.2cm]{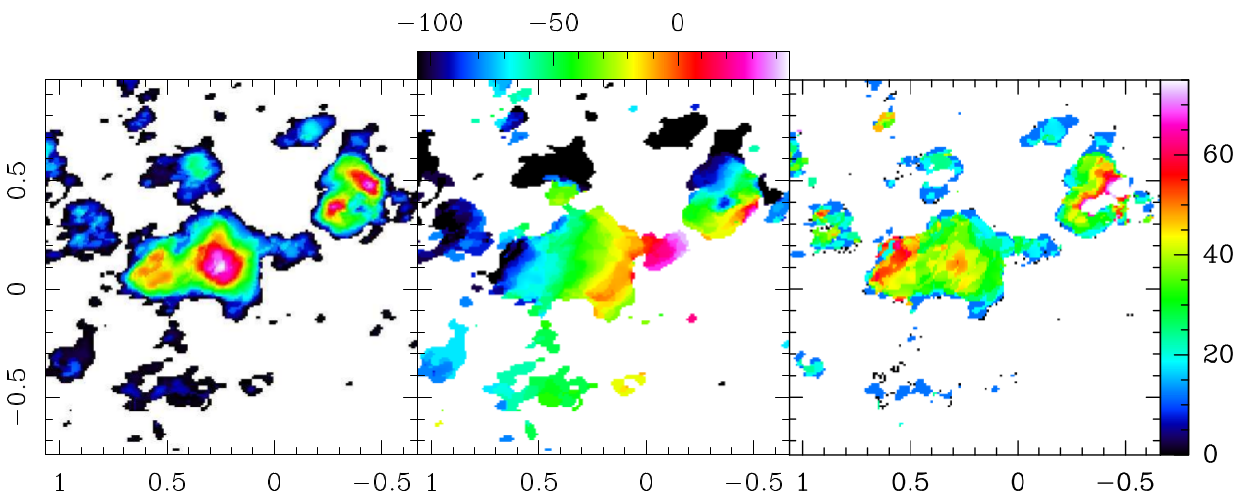}
}
\caption{ Same as Fig. \ref{fig:mod-1672} for NGC~1326 (left is the model, right the observations).
  The cube corresponds to the best PV fit, with a BH mass of  6.5 x 10$^6$ M$_\odot$
  and an inclination of  60$^\circ$.
}
\label{fig:mod-1326}
\end{figure*}

\begin{figure*}[h!]
\centerline{
  \includegraphics[angle=0,width=16cm]{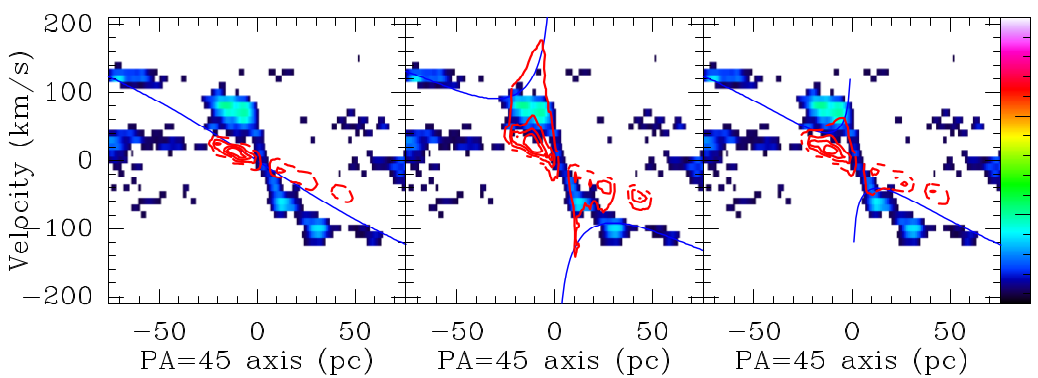}
}
\caption{Same as Fig. \ref{fig:3pv-1672} for NGC~1365,
without any black hole (left panel), with a black hole
of 6.9 x 10$^7$ M$_\odot$ from the M$_{\rm BH}$-$\sigma$ relation (middle panel), and
the best fit:  4 x 10$^6$ M$_\odot$ (right panel). The adopted inclination
i= 63$^\circ$ is from the galaxy to be conservative. Higher BH masss would
be derived with the lower  27$^\circ$ inclination estimated for the torus.
}
\label{fig:3pv-1365}
\end{figure*}

\begin{figure*}[h!]
\centerline{
  \includegraphics[angle=0,width=9.2cm]{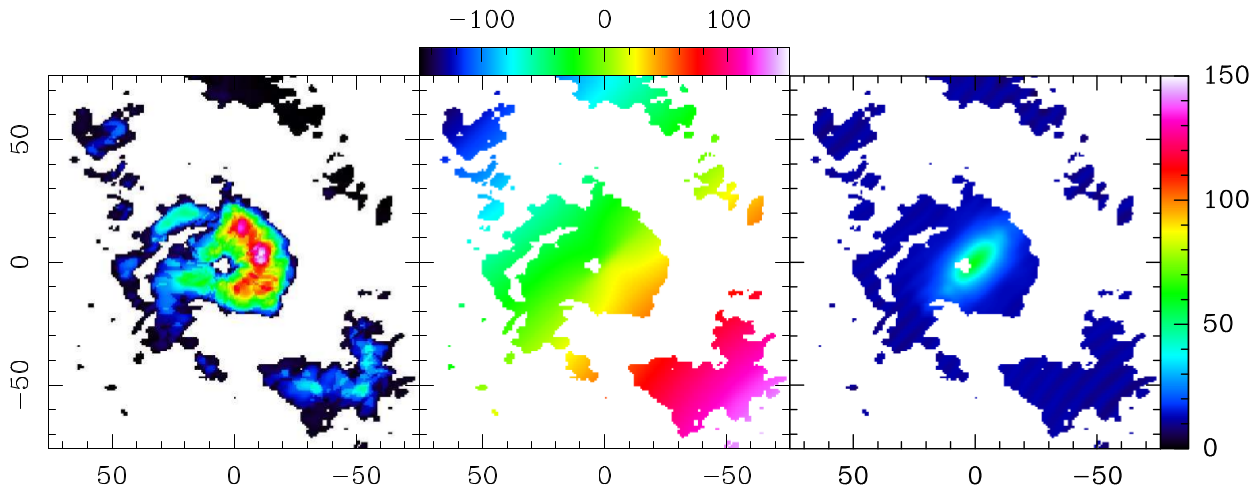}
    \includegraphics[angle=0,width=9.2cm]{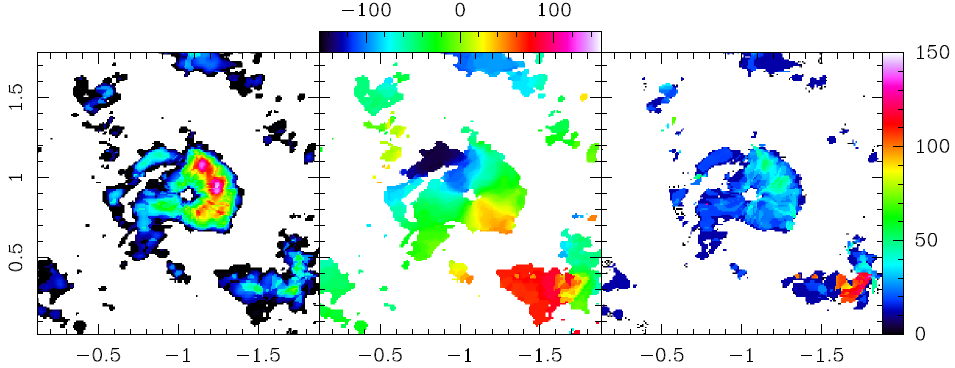}
}
\caption{ Same as Fig. \ref{fig:mod-1672} for NGC~1365 (left is the model, right the observations).
  The cube corresponds to the best PV fit, with a BH mass of  4 x 10$^6$ M$_\odot$
  and an inclination of  63$^\circ$.
}
\label{fig:mod-1365}
\end{figure*}

\begin{figure*}[h!]
\centerline{
  \includegraphics[angle=0,width=16cm]{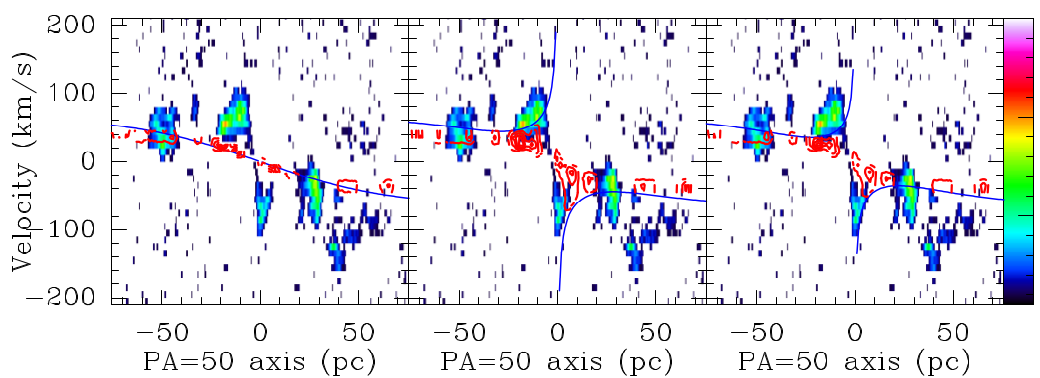}
}
\caption{Same as Fig. \ref{fig:3pv-1672} for NGC~1566,
without any black hole (left panel), with a black hole
of 1.35 x 10$^7$ M$_\odot$ from the M$_{\rm BH}$-$\sigma$ relation (middle panel), and
the best fit: 6.7 x 10$^6$ M$_\odot$ (right panel). As for  NGC~1365, the adopted inclination
i= 48$^\circ$  is from the galaxy.}
\label{fig:3pv-1566}
\end{figure*}

\begin{figure*}[h!]
\centerline{
  \includegraphics[angle=0,width=9.2cm]{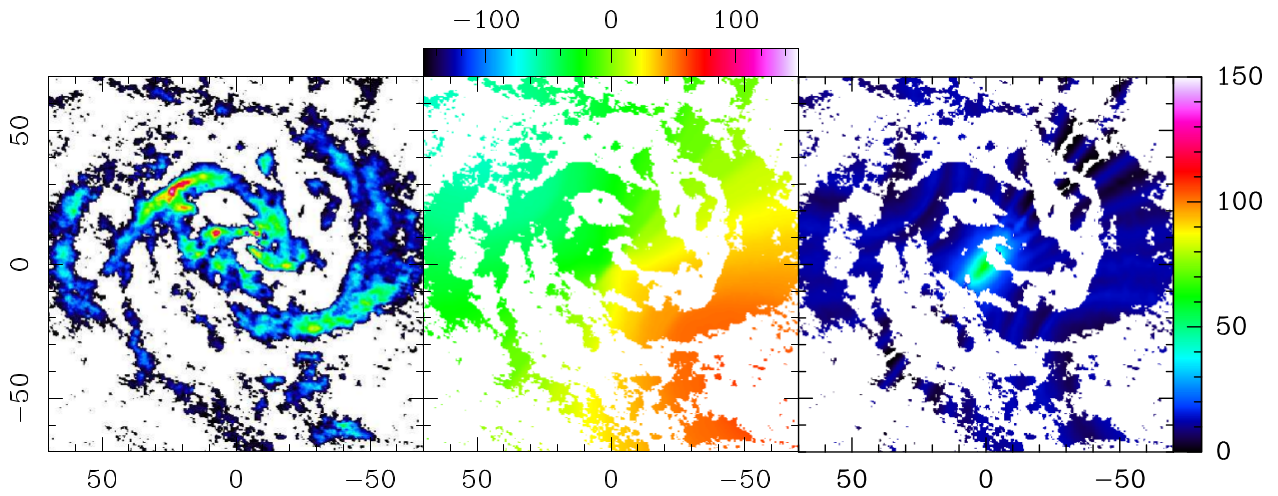}
    \includegraphics[angle=0,width=9.2cm]{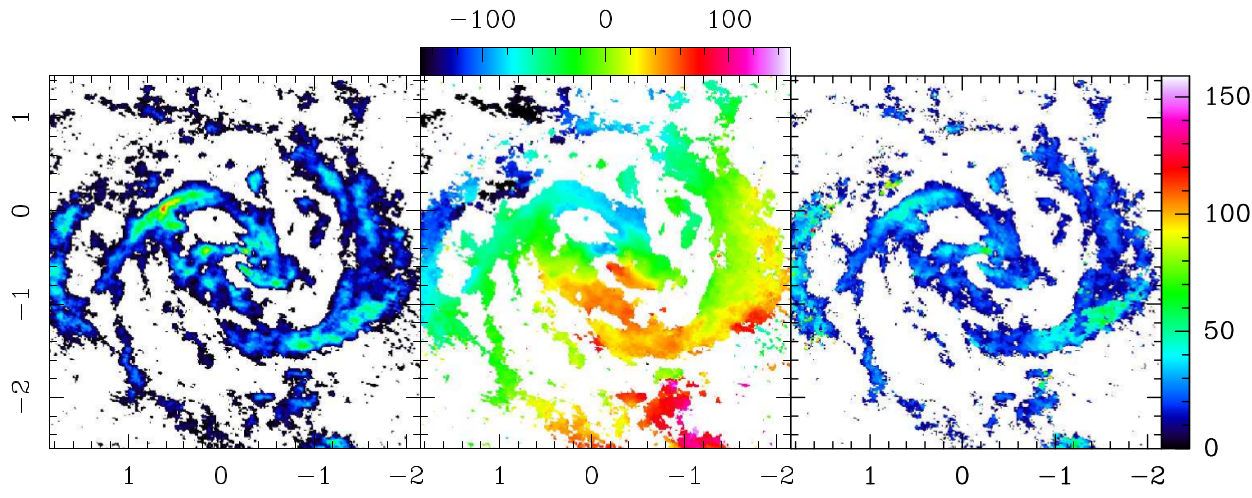}
}
\caption{ Same as Fig. \ref{fig:mod-1672} for NGC~1566 (left is the model, right the observations).
  The cube corresponds to the best PV fit, with a BH mass of  6.7 x 10$^6$ M$_\odot$
  and an inclination of  48$^\circ$.
}
\label{fig:mod-1566}
\end{figure*}

The same fitting procedure has been applied to NGC~613
(Figure \ref{fig:3pv-613} and \ref{fig:mod-613}, showing in addition
the perturbed velocity field and velocity dispersion due to the outflow),
NGC~1326 (Figs. \ref{fig:3pv-1326} and \ref{fig:mod-1326}),
NGC~1365, where the best compromise was obtained
with the outer disk inclination 
(Figs. \ref{fig:3pv-1365} and \ref{fig:mod-1365}), and the
same for NGC~1566 (Figs. \ref{fig:3pv-1566} and \ref{fig:mod-1566}).
The morphology was too complex in NGC~1808 to obtain
a satisfying fit, as mentioned in the previous section,
perhaps because of supernovae feedback molecular flows there.


\subsubsection{Summary of results on black hole masses}
\label{sec:BHmass}
The black hole masses  obtained are between 4 x 10$^6$ and 5 x  10$^7$ M$_\odot$,
in good agreement with the previous estimations of Tables \ref{tab:soi} and \ref{tab:bh2}:
they have the tendency to follow the relation obtained for pseudo-bulge galaxies, determined by
\cite{Ho2014}. It is now
fairly well established that classical bulges and ellipticals follow a tighter and shallower 
 M$_{\rm BH}$-$\sigma$ relation than
galaxies with pseudo-bulges (\citealt[e.g.,][]{Graham2011},
but see \citealt{Bennert2015}).
The eight galaxies studied in this paper can all be considered  pseudo-bulges:
the Sersic index of their bulges are all close to 1, and always $<$ 1.4 (Table 
\ref{tab:s4g}). Also their bulge-to-disk ratios are all below 0.4, which is
the location of pseudo-bulges \citep[e.g.,][]{Fisher2008, Gadotti2009}.

With the range of black hole masses found, the 
corresponding Eddington luminosities L$_{Edd}$ range from 
1.3 x  10$^{11}$ and 1.6 10$^{12}$ L$_\odot$, or log (L$_{Edd}$) in erg/s
between 44.7 and 45.8.
From the X-ray luminosities listed in  Table \ref{tab:sample} and
the bolometric corrections computed by \cite{Marconi2004}, we  
estimated the AGN bolometric luminosities
for all our galaxies, L$_{AGN}$ (see Table \ref{tab:bh2}).
The Eddington ratio L$_{AGN}$/L$_{Edd}$  in our galaxies is 
therefore mostly $<<$ 1, namely
between $\sim$0.2 (NGC 1068) and 3 $10^{-7}$ (NGC 1672).

\begin{figure*}[h!]
\centerline{
  \includegraphics[angle=-0,width=15cm]{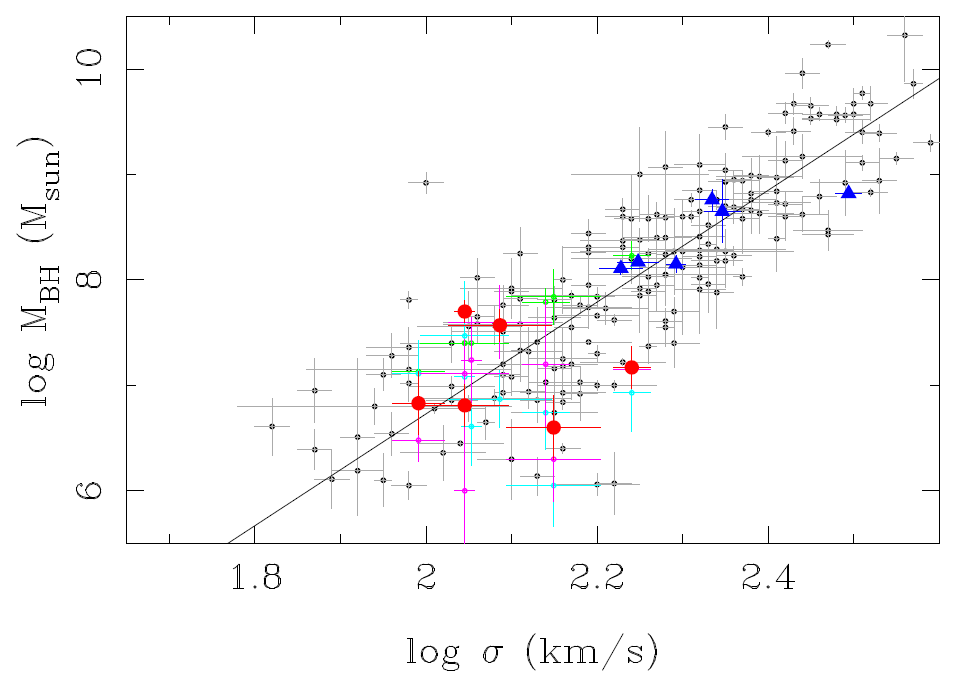}
}
\caption{Location of our galaxies in the M$_{\rm BH}$-$\sigma$ relation, compiled by
  \cite{vandenBosch2016}, represented by the gray points and the fitted line, of slope 5.35. Our galaxies are
  in red, while in blue are plotted other values recently obtained from the CO line:
  NGC~1097 \citep{Onishi2015}, NGC~1332 \citep{Barth2016}, NGC~3665 \citep{Onishi2017},
NGC~4429 \citep{Davis2018}, NGC~4526 \citep{Davis2013}, NGC~4697 \citep{Davis2017}.
Previous estimations of the BH masses of our sample galaxies are
        also plotted in green (Table \ref{tab:soi}), turquoise and pink
        (Columns 2 and 3 in Table \ref{tab:bh2}).
}
\label{fig:plot-remco}
\end{figure*}

The positions of the galaxies studied here in the M-$\sigma$ diagram are
displayed in Figure \ref{fig:plot-remco}, together with the compilation
by \cite{vandenBosch2016}. The values measured recently with CO emission
by the WISDOM collaboration are also plotted in blue: they apply to classical
bulges with much higher black hole masses. Galaxies with pseudo-bulges and 
with lower masses are harder to determine, and their relation is affected
by a larger scatter. In addition, barred galaxies appear in general below
the standard M$_{\rm BH}$-$\sigma$ relation \citep[e.g.,][]{Graham2011}, which contributes
to the scatter. 
 The scatter in barred galaxies may be due to a varying
velocity dispersion, because the orientation
of the bar is random with respect to the line of nodes. In barred galaxies,
gas driven into the center by gravity torques may boost a nuclear starburst,
which produces a $\sigma$-drop \citep{Wozniak2003}. Also, in pseudo-bulge
galaxies, the classical bulge is in general too light, and the black hole
mass might be better compared to the total baryonic mass \citep{DavisB2018}.
All eight galaxies studied in this paper are barred, and their
scatter (Figure \ref{fig:plot-remco}) is not unexpected. It is, however, crucial to
obtain BH masses  in this low-mass end region to better understand the different 
processes for black hole growth, either through mergers (classical bulges) or secular 
evolution (pseudo-bulges).

It is interesting to compare the black hole masses derived in this paper with the previous
estimations: in Fig. \ref{fig:plot-remco}, the green points are the values
derived from the M$_{\rm BH}$-$\sigma$ relation of \cite{Kormendy2013} and displayed 
in Table \ref{tab:soi}. These values are slightly above the M$_{\rm BH}$-$\sigma$ relation compiled
by \cite{vandenBosch2016}. The turquoise points are the values from  Column 2 in Table
\ref{tab:bh2}, estimated from the spiral pitch angle \citep{Davis2014},
and the pink points are the values from  Column 3 in Table \ref{tab:bh2} (various references).
The red points appear less scattered.

\section{Discussion and summary}
\label{disc}

We have presented our first ALMA results for a sample of 7 active galaxies,
all low-luminosity AGN. 
The high spatial resolution allows to enter the sphere
of influence of the central black holes, and to reveal the small-scale circumnuclear
molecular structures that we associate with possible molecular tori.
 For 6 out of 7 galaxies, we indeed detect a nuclear disk,
 decoupled in morphology and dynamics from the main disk.
Such a structure is also detected in an eighth galaxy studied by our group, 
NGC~1068 \citep{Garcia-Burillo2016, Garcia-Burillo2018}.

\subsection{Characteristics  of the tori}

The decoupled molecular tori (Table \ref{tab:torus})
may correspond to the required obscuration for type 2 AGN, even if
they do not have  the expected inclination for a type 2 classification. They sometimes show a gas hole
or depletion in the center (NGC~1365, NGC~1566, NGC~1672) which make
them approach a doughnut morphology. Their sizes extend over a wide range,
from 6 to 27~pc, but their masses are less scattered from $\sim$
1 to 4 10$^7$ M$_\odot$. There is no relation to the total stellar
masses of the galaxies, which vary from 1 to 9 10$^{10}$ M$_\odot$
(Table \ref{tab:s4g}).
The mass of the gas in the molecular tori appears 
slightly anticorrelated to the AGN strength, as traced by the X-ray luminosity
(Table \ref{tab:sample}), although with large scatter.  More galaxies
are needed to build a relation.

\subsubsection{Orientation of the tori}

One of the striking results emerging of these ALMA observations is that the
gas in the molecular tori is rarely aligned with the main disk, but 
present different inclination and position angles. This decoupling
was already observed for instance in NGC~1068 \citep{Gratadour2015,Garcia-Burillo2016}.
Given the widely different scales, 3-30~pc for the 
nuclear disks and 1-10kpc for the main disk, the dynamical
timescales are largely different, from $\sim$ 1 to 300 Myr.
It is then natural that these scales may decouple. It is sufficient 
that the central disk accretes some gas infalling with some
angle to the plane for it to warp and change its orientation,
averaging its old angular momentum vector with the tilted
one from the newly accreted gas. The accretion could come either 
from the kpc scales or from stellar mass loss or feedback
from its own star formation \citep[e.g.,][]{Emsellem2015}.
Alternatively, the tori could be unstable to non-axisymmetric modes growing
 on a dynamical timescale \citep{Papaloizou1984}. 

\subsubsection{Asymmetries: offsets and lopsidedness}

In addition, the molecular tori display asymmetries
and off-centering. These circumnuclear disks are frequently located
inside a star-forming ring, coincident with the  ILR of the bar, but are not centered in the ring (for instance NGC~1326).
The distance between the  disk center (scale of a few hundred~pc) and the AGN or torus
center (which coincide) is displayed in Table \ref{tab:torus}. 
They are on the order of a few tens of~pc, i.e., $\sim$ 10\% of the scale of the 
ILR rings.
This lopsidedness implies that the black hole is also wandering 
with a small amplitude around the center of mass of the galaxy.
Several mechanisms for this off-centering and implied
BH oscillations have been reviewed by \cite{Jog2009}.
The off-centering and wandering, characteristic of a Keplerian potential,
may be explained by the eccentric 
instability proposed by \cite{Shu1990},
when the masses of the central object and the disk are comparable.
This mechanism occurs with the 
exchange of angular momentum between the central mass and the disk
\citep{Woodward1994}. 
It will be interesting to establish whether the stellar component also
displays  $m=1$ instabilities, as  the M31 nucleus does, for example
\cite[e.g.,][]{Bender2005}. 
These instabilities may help to fuel the AGN as predicted by \cite{Hopkins2012}.

\subsubsection{Cases of the nuclear spirals}

In two cases, NGC~613 and NGC~1566, the CO emission has revealed
a nuclear spiral, which has the distinction of being trailing,
like the large-scale one. Inside the nuclear ring at the ILR of the bar,
 a leading spiral is usually expected to develop transiently and 
generate positive torques, which drive the inner gas onto the ring.
However, when the gravitational impact of the black hole is significant,
the spiral can then be trailing and the torques negative, to fuel
the nucleus \cite[e.g.,][]{Buta1996}. The special case of NGC~613
will be studied in a forthcoming paper \citep{Audibert2018}.

\subsection{Black hole masses}

The estimation of the black hole masses from the gas
kinematics suffers from specific uncertainties,   discussed in detail by previous authors
\citep{Barth2016,Onishi2015,Onishi2017,Davis2017,Davis2018}.
They discuss in particular the problem of edge-on systems,
which do not have  enough resolution on the minor axis or 
spatial resolution, which is barely equal to the radius of the SoI of the black hole. Our galaxies do not suffer from either
of these problems:
the spatial resolution of our ALMA observations is sufficient to
probe the SoI, and therefore the impact of the stellar potential,
and the uncertainties on the stellar mass-to-light ratio are less
important here.
However, our BH mass estimates still have the uncertainties
related to the distance scale, and the trade-off
needed to accommodate both the PV diagram fits and the observed
velocity dispersion maps as discussed below.

All our galaxies are barred, at various strength levels, and certainly
some non-circular motions are present, and must be included in
the uncertainties. A typical S-shaped velocity pattern is seen in the NGC~1672
circumnuclear region (Figure \ref{fig:mod-1672}). We postpone to
further work the simulation of the non-circular motions in these barred galaxies, including
both small and large scale.

Another caveat might come from the observed CO velocity dispersion: 
we tried to take into account this projected 
velocity dispersion when comparing the moment 2 of the data cubes
with the moments of our models.
However, the molecular gas is patchy, and especially inside 50~pc in radius
it is likely that there is an insufficient number of molecular clouds to sample
all the velocity gradients of the inner regions.
In the model with one million particles, we sample almost
continuously the velocity gradient, and this results in 
a large apparent dispersion along the line of sight towards the center in the case
of a massive black hole. This apparent dispersion is only due to 
a beam-smearing of the velocity gradient along the line of sight.
The intrinsic dispersion of the gas is negligible, between 2 and 10km/s.
It is even more negligible when there is a massive central mass, since
then the value of $\kappa$ the epicyclic frequency is significantly elevated,
and the critical velocity to stabilize  the disk in the Toomre sense is
very low. In some cases, even when the velocity gradient is high
and can only be obtained with a massive central component, the observed
dispersion is surprisingly low. There are cases where the molecular
gas is depleted or absent in the center, which is also a factor
that supresses the large dispersion. This has to be 
taken into account in the minimizing criteria.

Perhaps the most important uncertainty in the black hole mass determination
of these low-mass late-type objects is the decoupling of the molecular
torus. On the one hand, these nuclear disks are dense enough to give kinematical
information and subsist in the SoI, but their true inclination and position 
angle might be quite different from those of the main disks. It thus requires 
more 3D data to be able to disentangle both the decoupled morphology
and the actual motions around the black hole.

In summary, ALMA at high resolution brings a wealth of new information
on the decoupled molecular tori near the black holes. In the present paper,
we described the properties of the decoupled molecular tori
and estimated the mass of the central black holes, when possible,
as displayed in Fig. \ref{fig:plot-remco}.
In future work, we will address the fueling efficiency through torque
computation, and the feedback efficiency by estimating the gas outflows
and their energetics in these low-luminosity AGN.

\begin{acknowledgements}
 We warmly thank the referee for the constructive comments and suggestions. 
The ALMA staff in Chile and ARC-people at IRAM are gratefully acknowledged for their
help in the data reduction. We particularly thank Philippe 
Salom\'e for his useful advice.
SGB is thankful for the support from Spanish grant AYA2016-76682-C3-2-P.
LKH acknowledges funding from the INAF PRIN-SKA 2017 program 1.05.01.88.04. 
This paper makes use of the following ALMA data: ADS/JAO.ALMA\#2015.0.00404.S,
and ADS/JAO.ALMA\#2016.0.00296.S. 
ALMA is a partnership of ESO (representing its member states), NSF (USA), and NINS (Japan), 
together with NRC (Canada) and NSC and ASIAA (Taiwan), in cooperation with the Republic of 
Chile. The Joint ALMA Observatory is operated by ESO, AUI/NRAO, and NAOJ.
The National Radio Astronomy Observatory is a facility of the National Science Foundation 
operated under cooperative agreement by Associated Universities, Inc.
We used observations made with the NASA/ESA Hubble Space Telescope, and obtained 
from the Hubble Legacy Archive, which is a collaboration between the Space Telescope 
Science Institute (STScI/NASA), the Space Telescope European Coordinating Facility 
(ST-ECF/ESA), and the Canadian Astronomy Data Centre (CADC/NRC/CSA). 
We made use of the NASA/IPAC Extragalactic Database (NED)
and of the HyperLeda database. This work was supported by the Programme National
Cosmology et Galaxies (PNCG) of CNRS/INSU with INP and IN2P3, co-
funded by CEA and CNES.
\end{acknowledgements}

\bibliographystyle{aa}
\bibliography{torus1}
\end{document}